\documentclass{aa}  

\usepackage{graphicx}

\usepackage{color}
\usepackage{txfonts}
\usepackage{comment}
\usepackage{soul}

\usepackage{amssymb}
\usepackage{mathrsfs}

\usepackage{listings}
 
\usepackage{amsmath,amsfonts} 
\usepackage{xcolor}
\usepackage{enumitem}
\usepackage{booktabs}
\usepackage[normalem]{ulem} 

\usepackage{threeparttable}
\usepackage{hyperref}
\hypersetup{colorlinks=true, allcolors=blue}
\usepackage[table]{colortbl}
\usepackage{soul}

\definecolor{darkorange}{rgb}{0.9,0.2,0.0}
\definecolor{darkviolet}{rgb}{0.54, 0.1, 0.63}
\definecolor{darkgreen}{RGB}{3, 102, 3}

\newcommand{\gxchp}[1]{\texttt{GalaxyChop}}
\begin{document} 

\title{Untangling Stellar Components of Galaxies: Evaluation of Dynamical Decomposition Methods in Simulated Galaxies with \texttt{GalaxyChop}}

   \author{
    Valeria A. Cristiani \inst{1,2,3}\thanks{E-mail: valeria.cristiani@unc.edu.ar},
    Mario G. Abadi \inst{1,2},
    Antonela Taverna \inst{1,2,8},
    Juan Cabral \inst{1,4},
    Federico Benelli \inst{5,6},
    Bruno S\'anchez \inst{7}\\
   }

   \institute{
        Instituto de Astronom\'ia Te\'orica y Experimental -  Observatorio Astron\'omico de C\'ordoba (IATE, UNC--CONICET),  C\'ordoba, Argentina.
        \and
             Observatorio Astron\'{o}mico de C\'{o}rdoba, Universidad Nacional de C\'{o}rdoba, Laprida 854, X5000BGR, C\'{o}rdoba, Argentina.
        \and
          	Facultad de Matem\'atica, Astronom\'{\i}a y F\'{\i}sica, Universidad Nacional de C\'ordoba (FaMAF--UNC). Bvd. Medina Allende s/n, Ciudad Universitaria, X5000HUA, C\'ordoba, Argentina.
        \and 
            Gerencia De Vinculaci\'on Tecnol\'ogica
            Comisi\'on Nacional de Actividades Espaciales (CONAE), Falda del Ca\~nete, C\'ordoba, Argentina.
        \and 
            Instituto de Investigaci\'on y Desarrollo en Ingenier\'ia de Procesos y Qu\'imica Aplicada (IPQA, CONICET–UNC), Av. Velez Sarsfield 1611, CP. 5016, C\'ordoba, Argentina.
        \and 
            Facultad de Ciencias Exactas, F\'isicas y Naturales, Universidad Nacional de C\'ordoba (FCEFyN-UNC), Av. Velez Sarsfield 1611, CP. 5016, C\'ordoba, Argentina.
        \and    
            Aix Marseille Univ, CNRS/IN2P3, CPPM, Marseille, France.
        \and    
            Instituto de Astronomia, Universidad Nacional Autonoma de Mexico, Apdo. Postal 106, Ensenada 22800, B.C., Mexico.
    }
            
\titlerunning{Untangling Stellar Components of Galaxies}
\authorrunning{Cristiani et al.} 
\date{Received September 11, 2024; accepted September 24, 2024}

  \abstract{Galaxy formation is intrinsically connected to the distinct evolutionary processes of disk and spheroidal systems, which are the fundamental stellar components of galaxies. Understanding the mutual dynamical interplay and co-evolution of these components requires a detailed dynamical analysis to allow for a disentanglement between them. We introduce JEHistogram, a new method for the dynamical decomposition of simulated galaxies into disk and spheroidal stellar components, utilizing the angular momentum and energy of star particles. We evaluate its performance against five previously established methods using a sample of equilibrium galaxies with stellar masses in the range $10^{10} \leq M_\mathrm{gal}/M_\odot \leq 10^{12}$. Our assessment involves several metrics, including the completeness and purity of stellar particle classification, scale lengths, mass density profiles, velocity dispersion, and rotational velocity profiles. While all methods approximate the properties of the original components, such as mass fractions and density or velocity profiles, JEHistogram demonstrates a better accuracy, particularly in the inner regions of galaxies where component overlap complicates separation. Additionally, we apply JEHistogram to a Milky Way-like galaxy from the IllustrisTNG cosmological simulations, showcasing its capability to derive properties like size, mass, velocity, color, and age of dynamically defined disk and spheroidal components. All dynamical decomposition methods analyzed are publicly accessible through the Python package \gxchp{}.}


   \maketitle
%
\section{Introduction}

Galaxies are complex self-gravitating stellar systems consisting mainly of stars, dark matter, gas, and dust. 
From these early catalogs \citep{Messier_1781, Herschel_1864, Dreyer_1888}, through Hubble \citep{Hubble_1926, Hubble_1936}, to the present day, the images show a morphological variety that reveals a great diversity of stellar components \citep{Sandage&Tammann_1981}.

The analysis of the galaxy light distribution allows us to make the most basic morphological distinction. By determining whether galaxies are dominated by the presence of a rotation supported disk or a velocity dispersion supported spheroid they can be categorized as late-type or early-type galaxies respectively. As instrumentation progress grew, this simple classification became quite insufficient as sub-structures gradually became revealed in galaxies, such as the thin and thick disk, the nucleus, the stellar halo, and the bar \citep{2002-Dalcanton,2004-Kormendy,2013-Trujillo}. These sub-structures, normally named stellar components interact with each other and also follow their own individual temporal evolution. Consequently, the description of the formation and evolution of galaxies is extremely related to the formation and evolution of their components and their coupling.

Cosmological N-body numerical simulations, like The EAGLE project \citep{Schaye-2015-EAGLE, Crain-2015-EAGLE} or The Next Generation Illustris Simulations (IllustrisTNG) \citep{Pillepich-2018-TNG, Naiman-2018-TNG, Nelson-2018-TNG, Marinacci-2018-TNG, Springel-2018-TNG}, have become one of the fundamental tools for studying galaxies and their evolution over time. The progress in the development and optimization of the implementation of galaxy formation and evolution models allowed us to increase the particle resolution and the box size of the simulations \citep{2020_Vogelsberger}. The possibility of having well-resolved galaxies with multiple substructures allows us to study their physical properties, assembly, and temporal evolution. Thus, to be able to select stellar particles in different components such as thin disk, thick disk, bulge, stellar halo bar, etc., the community has implemented different strategies to identify such substructures based on dynamic quantities. The pioneering work in carrying out this task was that of \cite{abadi_03}, who presented a method for selecting to which stellar component each stellar particle belonged. Due to its simplicity and efficiency, this dynamical decomposition technique has been widely used with some simplifications or modifications \citep[e.g.][]{Okamoto-2008, Scannapieco-2009, 2012_Tissera, Domenech+12, 2014_Vogelsberger, 2014_Marinacci, obreja_gsf_code, 2019_Park, du_19, Xu-2019, Gargiulo-2019, Jagvaral-2022, Zana-2022, Yu-2023}. However, there is no study or agreement on which implementation recovers the components more accurately for specific metrics, such as the completeness and purity of stellar particle classification or the ability to recover physical features like scale lengths, mass density profiles, velocity dispersion, and rotational velocity profiles. Authors usually choose among some of the variations of the method available in the literature or modify one of them to carry out their studies. This may depend on factors such as the number of stellar components to be identified or the computational cost required to perform the dynamical decomposition of the galaxies of interest. Therefore, it is useful to have the information provided by these different implementations, to compare the advantages and disadvantages of each one, in addition to the differences that arise from using one or the other. With all this aim in mind, in this work, we present the dynamical decomposition method named JEHistogram, which is an improvement of the one presented by \citet{abadi_03} that takes into account the energy distribution of the particles, in addition to the angular momentum distribution of the particles. Its implementation, as well as the other 5 dynamic decomposition methods: JThreshold, JHistogram, KMeans, GaussianMixture, and AutoGaussianMixture (see Sect.~\ref{subsec:JEH method} and \ref{subsec:models} for a brief description of each one), with which the analysis is carried out, is present in a software package called \gxchp{} that enables the dynamical decomposition of galaxies in N-body simulations. Besides, \gxchp{} can calculate dynamical quantities and obtain easily the results.

This paper is organized as follows: in Sect.~\ref{sec:dyn_dec} we briefly describe the theory behind the different methods of dynamical decomposition of simulated galaxies, as well as the methods implemented in the \texttt{GalaxyChop} package. The performance of the dynamical decomposition methods applied to model galaxies is in Sect.~\ref{sec:Performance}. In Sect.~\ref{sec:galaxychop} we dynamically decompose a galaxy belonging to the IllustrisTNG simulations and analyze the results obtained with the different methods. Finally, in Sect.~\ref{sec:conclusion}, we summarize our results and conclusions.

\section{Dynamic decomposition}\label{sec:dyn_dec}

\begin{figure}
	\includegraphics[width=\columnwidth]{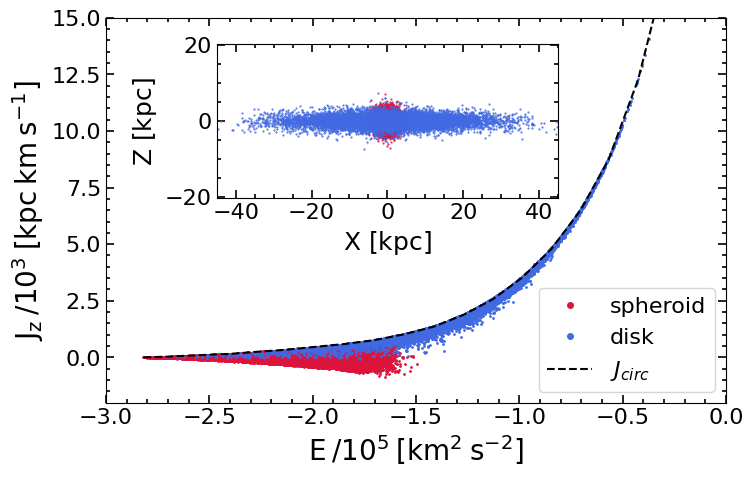}
    \caption{Lindblad diagram of a model galaxy in equilibrium (see galaxy model 04 in Table~\ref{tab:masas_y_f_sph}). The z component of the angular momentum is shown as a function of the specific binding energy for all stellar particles. In red are represented the particles corresponding to the spheroid while in blue are represented the stellar particles of the disk. The black dashed line is the $J_\mathrm{circ}$. The top inset shows the edge-on view of the galaxy to which the diagram belongs.}
    \label{fig:limblad}
\end{figure}

A simple way to describe the dynamic properties of the particles inside an isolated system in equilibrium is to use the integrals of motion since they do not vary as a function of time. An example of this is the Lindblad diagram \citep{Lindblad_1933}, which is constructed based on the energy (Eq.~\ref{eq:eq_energy}) and angular momentum (Eq.~\ref{eq:eq_jz}) integrals. Assuming a galaxy as an isolated system in equilibrium with axial symmetry, the orbit of the i-th particle is a fixed point in the Lindblad diagram and can therefore be characterized by its specific total energy 
\begin{equation}\label{eq:eq_energy}
    E_i = \frac{v_i^2}{2} - \sum_{j \neq i} \frac{G m_j}{r_{ij}}
\end{equation}
where $v_i$ is their total velocity, $m_j$ is the mass of the rest of the particles in the galaxy, and $r_{ij}$ is the distance between the i-th and j-th particle. The other quantities that characterize the orbit of the particle in the Lindblad Diagram are the z-component of the specific angular momentum of the i-th particle
\begin{equation}\label{eq:eq_jz}
    J_{z,i} = (x v_y - y v_x)_i
\end{equation}
where $x, y$ correspond to the x and y position and  $v_{x}, v_{y}$ are the velocities in the x and y directions. Note that it is assumed that the z-axis coincides with the spin axis of the galaxy. Figure~\ref{fig:limblad} shows the Lindblad diagram for the star particles of a simulated galaxy model. The stellar particles in red belong to the spheroid while the blue ones belong to the disk.

In the case of galaxies, an important aspect is that the positions of the particles in the Lindblad diagram are energetically bounded by $E <$ 0 since it is a gravitationally bound system. Also, the inequality $-J_\mathrm{circ} \leq J_z \leq J_\mathrm{circ}$, where $J_\mathrm{circ}$ corresponds to the specific angular momentum of a circular orbit for a given energy value \citep{van_den_Bosch_1999}, provides an upper and lower bound on the region that orbits can populate at in angular momentum values. This is because given an energy value E, the maximum angular momentum value $J_\mathrm{circ}$ corresponds to the angular momentum of a star particle in a co-rotating circular orbit, while the minimum angular momentum value $-J_\mathrm{circ}$ corresponds to the angular momentum of a star particle in a counter-rotating circular orbit. Therefore, the Lindblad diagram is a tool that allows a detailed analysis of the components of the galaxy through a statistical study of the dynamical properties of the particles that constitute it; since the particles belonging to different components populate different areas of the diagram. Thus, the particles that belong to the disk, which have circular orbits, have values of $J_{z}\sim J_\mathrm{circ}$; while the particles that belong to the spheroid have $J_{z}\sim$ 0, as can be seen in the Fig.~\ref{fig:limblad}. The ratio between the z-component of the angular momentum $J_z$ and the circular angular momentum $J_\mathrm{circ}$ of the stellar particles for a given value of energy ($\epsilon = J_z(E)/J_\mathrm{circ}(E)$) is called {\it orbital circularity or circularity parameter} \citep{van_den_Bosch_1999}. This parameter makes it possible to distinguish in the first instance two stellar substructures in a galaxy since stellar particles belonging to the disk component take values of $\epsilon\sim$ 1, while those that form the spheroid are around $\epsilon\sim$ 0.

The analysis of particle clustering in the dynamical space is a powerful tool for classifying stars as belonging to different stellar components. Therefore, most decomposition models use methods that combine some or all of the properties that comprise the dynamic space. With the aim of untangling the stellar components of galaxies, in the following, we present a new decomposition method, JEHistogram, along with five other decomposition models developed by different authors. All six methods have been implemented in the Python package \texttt{GalaxyChop}\footnote{\href{https://github.com/vcristiani/galaxy-chop}{https://github.com/vcristiani/galaxy-chop}} which will be used throughout the rest of the paper for comparison.

\subsection{JEHistogram method}\label{subsec:JEH method}

Our method improves upon the original implementation by \citet{abadi_03}. We summarize the main assumptions here and refer readers to that paper for further details. The fundamental assumption is that the spheroidal component has zero net angular momentum. All counter-rotating particles -those with a negative z-component of angular momentum (where the positive z-axis is defined by the total angular momentum of all stellar particles)- belong to the spheroidal component. To counterbalance these counter-rotating stars, we add an equal number of co-rotating particles, randomly selected from those with a positive z-component of angular momentum. These co-rotating particles have a circularity parameter $\epsilon=J_z/J_\mathrm{circ}$ following the same distribution but with a positive rather than negative sign. This approach essentially builds up a spheroidal component with a circularity histogram that is symmetric around $\epsilon = 0$. All remaining galaxy particles are co-rotating and assigned to the disk component, ensuring that no counter-rotating particles are assigned to the disk (see \citet{abadi_03}, Fig.~2, inset lower panel). Note that in the original \citet{abadi_03} method, there are no restrictions on the binding energy of the particles, so co-rotating particles in the spheroid can have different energies compared to the counter-rotating ones. To correct this asymmetry, we randomly select spheroidal co-rotating particles from all co-rotating stars, ensuring that their binding energy distribution is similar to that of the counter-rotating particles. As a result, the stellar particles belonging to the spheroidal component have an (almost) symmetric circularity distribution, peaked around $\epsilon = 0$, but co-rotating and counter-rotating stars having very similar binding energies. Technically, we build a grid in the $\epsilon=J_z/J_\mathrm{circ}$ versus $E/|E|_\mathrm{max}$ space, where $|E|_\mathrm{max}$ is the binding energy of the most bound particle, thus instead of working with the original Lindblad diagram (see Fig.~\ref{fig:limblad}) we work with the normalized Lindblad diagram (see as an example Fig.~\ref{fig:resultado_en_el_espacio_de_parametros}) to make it more straightforward to apply to galaxies of a wide range of masses. To decide which co-rotating stellar particles are assigned to the spheroid we select from each two-dimensional bin with $\epsilon=J_z/J_\mathrm{circ} > 0$ the same number of particles that has the bin symmetric to it, with respect to $\epsilon=J_z/J_\mathrm{circ} = 0$. Finally, all stellar particles not assigned to the spheroid are classified as part of the disk. In this case we have used 100 bins in $\epsilon=J_z/J_\mathrm{circ}$ and 20 bins in $E/|E|_\mathrm{max}$.

\subsection{Other methods of dynamical decomposition}\label{subsec:models}

We briefly describe below the other 5 different ways of assigning particles to one or another component, implemented in the \texttt{GalaxyChop} package.

\begin{table*}
    \caption{Total stellar mass, spheroid mass, disk mass, spheroid mass fraction, spheroid particle number and disk particle number of the 9 models of galaxies in equilibrium using the {\sc AGAMA} package \citep{agama}.}
        \label{tab:masas_y_f_sph}
    $$
        \begin{array}{m{0.2\linewidth} rrrrrr}
            \hline
            \noalign{\smallskip}
            AGAMA & M_{gal} & M_{sph} & M_{dsk} & f_{sph} & N_{sph} & N_{dsk}\\
            Model\ Number & [10^{10} M_\odot] & [10^{10} M_\odot] & [10^{10} M_\odot] & & & \\
            \noalign{\smallskip}
            \hline
            01  & 1.00   &  0.50  & 0.50  & 0.50 & 3\,125   & 3\,125   \\
            02  & 1.78   &  0.97  & 0.80  & 0.54 & 6\,113   & 5\,001   \\ 
            03  & 3.16   &  1.90  & 1.26  & 0.60 & 11\,859  & 7\,906   \\ 
            04  & 5.62   &  3.65  & 1.92  & 0.65 & 22\,845  & 12\,301  \\ 
            05  & 10.00  &  7.00  & 2.66  & 0.70 & 43\,750  & 18\,750  \\ 
            06  & 17.78  &  13.65 & 4.27  & 0.77 & 83\,357  & 27\,786  \\ 
            07  & 31.62  &  25.30 & 6.31  & 0.80 & 158\,114 & 39\,528  \\ 
            08  & 56.23  &  47.73 & 8.40  & 0.85 & 298\,744 & 52\,719  \\ 
            09  & 100.00 &  90.00 & 10.00 & 0.90 & 562\,500 & 62\,500  \\ 
            \noalign{\smallskip}
            \hline
        \end{array}
    $$
\end{table*}

\begin{itemize}
    \item JThreshold: This method is the simplest model used to decompose galaxies. It is an implementation that assigns stellar particles to a spheroid or disk component using only the circularity parameter $\epsilon$. A threshold ($\epsilon_\mathrm{cut}$) is defined by the user and the particles with $\epsilon > \epsilon_\mathrm{cut}$ are assigned to the disk and the particles with $\epsilon \leq \epsilon_\mathrm{cut}$ are assigned to the spheroid component. Several authors use different values of $\epsilon_\mathrm{cut}$, such as \citet{2014_Vogelsberger, 2014_Marinacci, 2019_Park}. This model is usually implemented to classify galaxies by morphology.

    \item JHistogram: This method is the implementation of the model described by \citet{abadi_03}. As in the two previous cases, this method decomposes the galaxy into two components: disk and non-rotating spheroid. It is assumed that the distribution of the $\epsilon$ for the spheroid component is symmetric around $\epsilon=$ 0. For this purpose, the distribution of $\epsilon<$ 0 is taken and a distribution of $\epsilon>$ 0 reflected around $\epsilon=$ 0 is constructed, randomly selecting stellar particles with $\epsilon>$ 0. The rest of the particles are assigned to the disk component.

    \item KMeans: This method is an implementation of the \texttt{K-means} package of \texttt{scikit-learn} \citep{scikit-learn} library. The clustering model of \texttt{K-means} is a no-supervising machine learning technique. In this implementation, the algorithm looks for a specific number of clusters ($k$) inside a three-dimensional space formed by $e/|e|_\mathrm{max}$, $\epsilon$ and $\epsilon_\mathrm{p}$, where $e$ is the specific binding energy, $|e|_\mathrm{max}$
    is the specific binding energy of the most bound particle and
    $\epsilon_\mathrm{p} = J_\mathrm{proj}(E)/J_\mathrm{circ}(E)$ with $J_\mathrm{proj}~=~J-J_z$ is the projected angular momentum. These $k$ clusters are interpreted as different stellar components of the galaxy.
    
    \item GaussianMixture: This case is an implementation of the model developed by \citet{obreja_gsf_code}. The implementation of this model applies the Gaussian Mixture Model \citep{scikit-learn} algorithm to identify the different clusters in the three-dimensional space formed by $e$, $\epsilon$ and $\epsilon_\mathrm{p}$. In the same way as Kmeans, the number of Gaussians used for the dynamic decomposition is associated with the number of stellar components of the galaxy to be identified. This method associates to each stellar particle a probability of belonging to each Gaussian to be identified.

    \item AutoGaussianMixture: This is an implementation of the model developed by \citet{du_19}. This model, as the previous one, uses the Gaussian Mixture Model in the parameter space ($e$, $\epsilon$, $\epsilon_\mathrm{p}$). The number of Gaussian is automatically selected using a variation of the Bayesian Information Criterion (BIC) \citep{1978_Schwarz}. Each component is associated with a physical structure taking into account the mean values of $e$, $\epsilon$, $\epsilon_\mathrm{p}$ of each Gaussian. In this case, as in the methods described above, the three-dimensional parameter space over which the dynamic decomposition is carried out is ($e$, $\epsilon$, and $\epsilon_\mathrm{p}$). To separate the disk component from the spheroid one, it uses $\epsilon=$ 0.5 as the threshold. The disk is subdivided into thick and thin disk using the threshold $\epsilon=$ 0.85. The spheroidal component is subdivided into halo and bulge, using, in this case, a threshold on specific normalized energy $e=-$ 0.75. Note that the specific energy is normalized to the value of the specific energy of the most bound particle. Like the \texttt{GaussianMixture} method, it returns for each stellar particle the probability of belonging to each of the components of the galaxy.
\end{itemize}

Note that the specific computation time of each method (except AutoGaussianMixture) is practically negligible ($\sim 10^{-2} - 10^{-1}$) compared to the CPU time invested in the potential energy calculations. Since our implementation of the potential energy calculation is parallelized, the dependence on particle number is much better than theoretical $N^2$ and scales approximately as $t_\mathrm{CPU} \propto N^{0.75}$ on average.

\section{Performance of dynamical decomposition models}\label{sec:Performance}

To understand the differences obtained in the identification of components by the 6 variations of the dynamical decomposition, described in Sect.~\ref{subsec:JEH method} and~\ref{subsec:models}, we performed an analysis of the assignment of stellar particles to each component. For this purpose, we build a set of 9 models of galaxies in equilibrium with a discoidal stellar component and a spheroidal one using the AGAMA package \citep{agama}. Then we assign the stellar particles to the disk and spheroidal components through dynamic decomposition methods and compare some of the physical properties of the resulting components with the original ones.

\subsection{Sample of model galaxies in equilibrium}\label{subsec:muestra}

Each of the 9 models of galaxies in equilibrium of the sample has a spheroidal stellar component and a discoidal stellar component. They were generated using the {\sc AGAMA} code \citep{agama}, which allows us to simulate isolated galaxies in equilibrium from ad-hoc initial conditions. The range of stellar masses spanned by the sample is $\rm{10 < \log_{10}(M_{gal}/M_\odot) < 12}$ and the range of spheroidal mass fractions is $f_\mathrm{sph} = 0.5-0.9$. Table~\ref{tab:masas_y_f_sph} shows the stellar masses of galaxies, of their disk and spheroid components, the mass fraction of the spheroid, and the number of particles in each component. Furthermore, in all cases, the stellar mass density profile of the discoidal component corresponds to an exponential profile. In contrast, the mass density profile of the spheroid corresponds to a power law with an exponential decay (see Table~1 of \citet{AGAMA-docu}). 

\begin{figure}
    \includegraphics[width=\columnwidth]{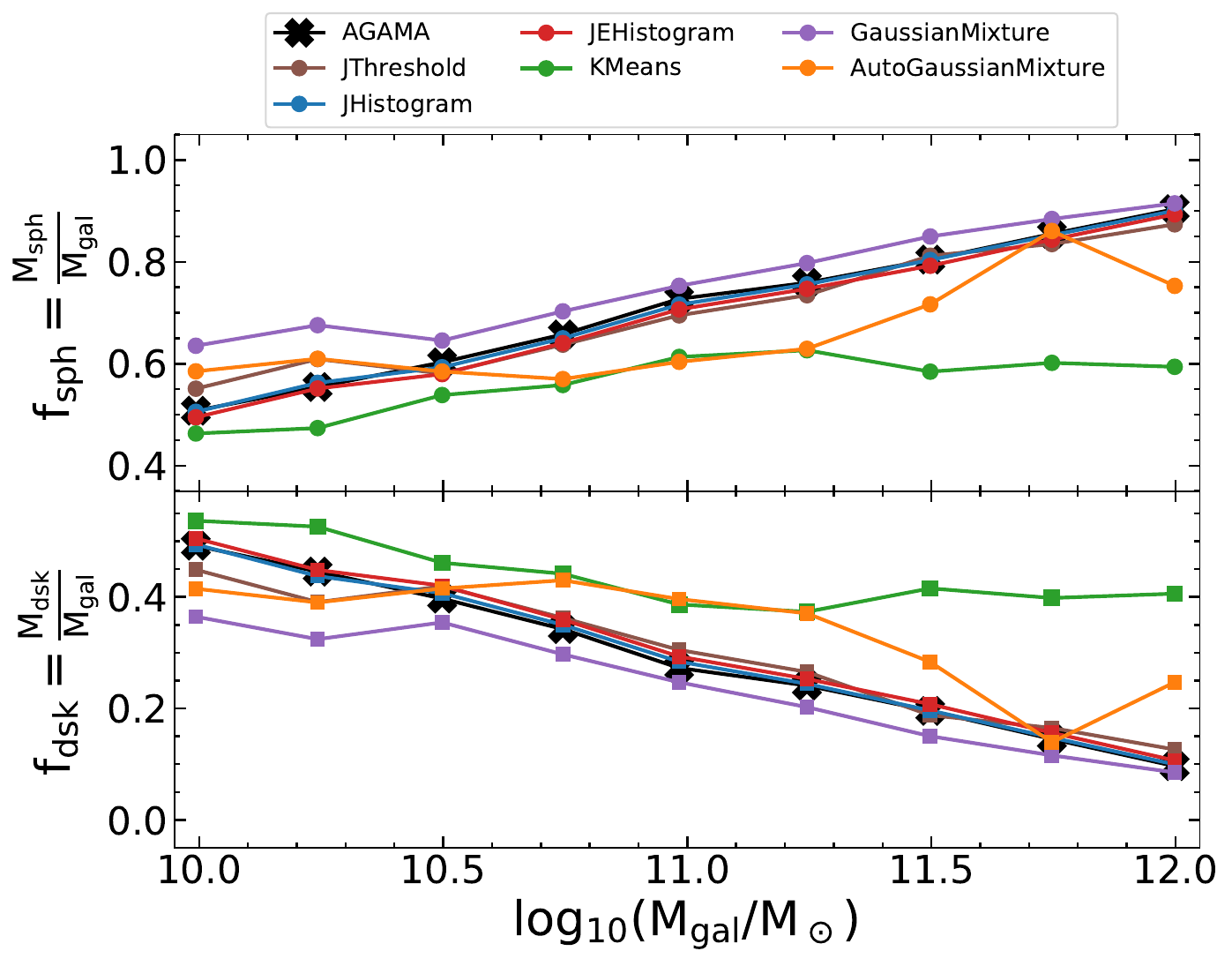}
    \caption{Upper panel: Mass fractions of the spheroidal component $f_\mathrm{sph}$ as a function of galaxy stellar mass $\rm{M_{gal}}$. Lower panel: Mass fractions of the discoidal component $f_\mathrm{dsk}$ as a function of galaxy stellar mass $\rm{M_{gal}}$. In both cases, the crosses with black solid line represent the original $f_\mathrm{sph}$ or $f_\mathrm{dsk}$ values for the 9 equilibrium galaxies built using the {\sc AGAMA} code \citep{agama} and the circles with a solid line show the $f_\mathrm{sph}$ or $f_\mathrm{dsk}$ values recovered with each of the 6 dynamical decomposition methods used.}
    \label{fig:fracciones_de_masa_de_galaxias_modelo}
\end{figure}

\subsection{Stellar component mass fraction}\label{subsec:fracc_de_masa}

\begin{figure*}
	\includegraphics[width=\textwidth]{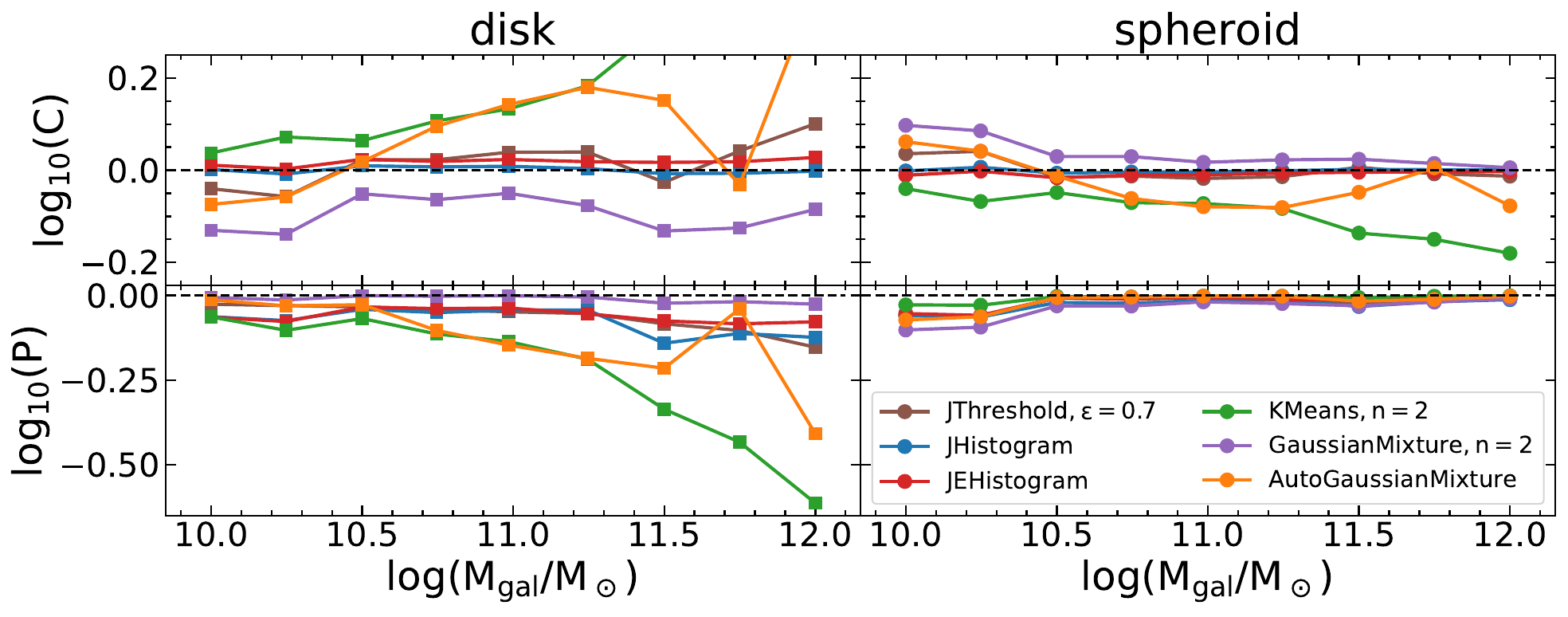}
    \caption{Completeness $C$ and purity $P$ of the recovered components in the sample of 9 equilibrium galaxy models, with each of the 6 methods of dynamical decomposition. The upper panels correspond to the completeness of the disks (left) and the spheroids (right), while the lower panels correspond to the purity of the disks (left) and the spheroids (right). The dashed black lines in the upper panels represent a completeness of 1, indicating that the number of particles assigned to the component by the method matches the actual number of particles in the component. The dashed black lines in the lower panels represent a purity of 1, indicating that the number of particles assigned to each component was correctly assigned.}
    \label{fig:pureza_y_completitud}
\end{figure*}

The zeroth-order test compares the stellar component mass fractions that each method reports, concerning the original stellar component mass fractions for each galaxy. In Fig.~\ref{fig:fracciones_de_masa_de_galaxias_modelo} we present the spheroid mass fraction $f_\mathrm{sph}$ (upper panel) and disk mass fraction $f_\mathrm{dsk}$ (lower panel), as a function of the $\rm{M_{gal}}$ for the 9 equilibrium galaxy models and the 6 implementations of dynamic decomposition methods. In addition, we present the spheroid mass fractions for observational data for reference. As shown in the upper panel of Fig.~\ref{fig:fracciones_de_masa_de_galaxias_modelo} most of the 6 methods can reproduce the spheroidal mass with differences relative to the original mass fraction between 0.1\% to 17.4\%. The JHistogram and JEHistogram methods achieve the best performance for this stellar component since the differences become even smaller than 3.8\%. The differences to $f_\mathrm{sph}$ for the method JThreshold reach up to 10\% while AutoGaussianMixture achieves up to 17\%. The KMeans method underestimates the $f_\mathrm{sph}$ regardless of the stellar mass of the galaxy, reaching the largest differences, above 27\%, for the 3 most massive galaxies. On the contrary, it is observed that the GaussianMixture method overestimates the mass fractions in all cases, exceeding differences of 21\% for the two less massive galaxies.

It is interesting to know how the methods recover the disk and spheroid mass fractions, and how these fractions depend on the numerical resolution at which the galaxy is simulated for a given method. We used our fiducial galaxy model (AGAMA-04, see Table \ref{tab:masas_y_f_sph} and Fig.~\ref{fig:limblad}) represented by $N\sim35\,000$ stellar particles, and we tested whether the recovered mass fraction $f_\mathrm{sph}$ depends on $N$. We decreased and increased $N$ to $N\sim3\,500$ and $N\sim350\,000$, respectively, and found that $f_\mathrm{sph}$ changes from 59\% to 65\%, compared to 64\% for $N\sim35\,000$. Note that for the highest resolution tested, $N\sim350\,000$, $f_\mathrm{sph}$=65\% has already converged to the original mass fraction with which the model was built, see black curve labeled AGAMA in upper panel of the Fig.~\ref{fig:fracciones_de_masa_de_galaxias_modelo}. The JHistogram method shows similar convergence behavior, though it is slightly more accurate, while the JThreshold, KMeans, and GaussianMixture methods do not show any convergence, only small fluctuations of the order of 1\%.

As the mass fraction of the disk is $1-f_\mathrm{sph}$, the trends are reversed. So those methods that overestimate the spheroid underestimate the disk and vice versa (see bottom panel of the Fig.~\ref{fig:fracciones_de_masa_de_galaxias_modelo}).

Although the methods can retrieve the stellar mass fractions of the components in most cases, this does not guarantee that the particles have been assigned to the correct component.

\begin{figure*}
	\includegraphics[width=\textwidth]{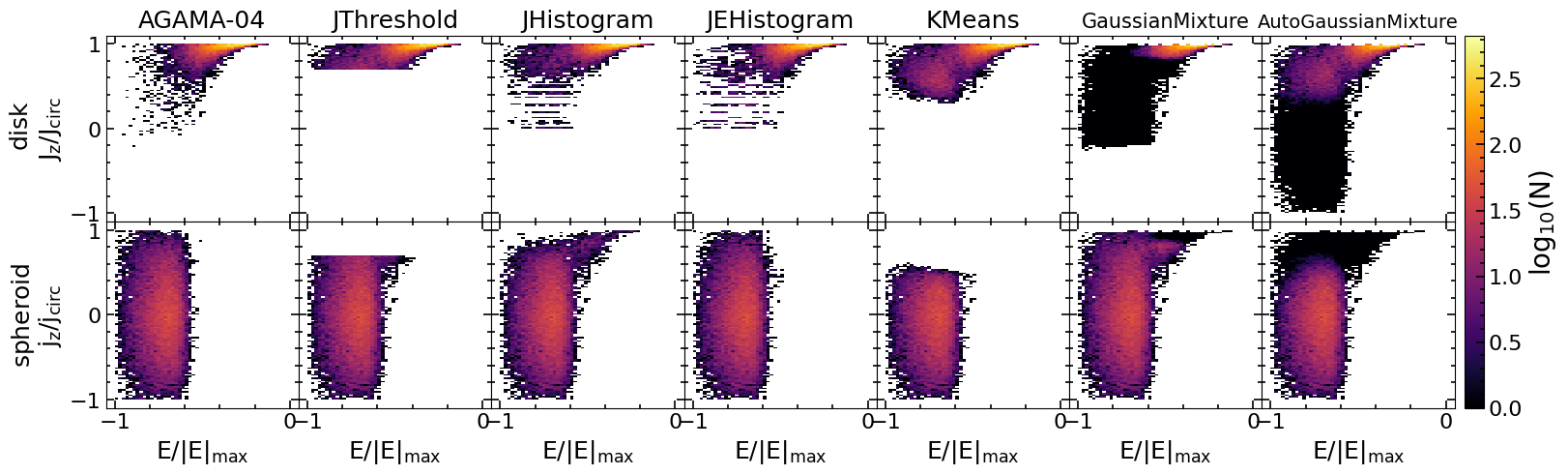}
    \caption{Distribution of stellar particles of the disk (first row) and spheroid (second row) of the AGAMA-04 galaxy in the normalized Lindblad diagram. The first column corresponds to the stellar particle distribution of the original disk and spheroid, while the remaining columns correspond to that obtained by each of the 6 methods. The color code represents the number of stellar particles in each region.}
    \label{fig:resultado_en_el_espacio_de_parametros}
\end{figure*}

\subsection{Purity and Completeness}\label{subsec:pyc}

One way to quantify how successful dynamic decomposition methods are is through parameter completeness and purity. We define completeness $C$ as
\begin{equation}
    C = \frac{N_{dyn}}{N_{AGAMA}},
\end{equation}
where $N_{dyn}$ is the number of particles assigned to the component by the method and $N_{AGAMA}$ is the actual number of particles in the original component. Since the mass of the stellar particles is the same in both components, in the 9 equilibrium galaxy models, we can interpret this quantity as the ratio between the mass fraction of the component identified with the dynamical decomposition method ($f_\mathrm{sph}$ o $f_\mathrm{dsk}$) and the mass fraction of the original component. In Fig.~\ref{fig:fracciones_de_masa_de_galaxias_modelo} this corresponds to, in the upper panel, the ratio of colored dots to black crosses, and in the lower panel, the ratio of colored squares to black crosses. However, recovering the correct number of particles per component does not guarantee that these have been correctly assigned to the corresponding stellar component. Additionally, we define the purity $P$ as
\begin{equation}
    P = \frac{N_{dyn,\ true}}{N_{dyn}},
\end{equation}
that is the fraction between the number of particles correctly assigned to the component $N_{dyn,\ true}$ and the number of particles assigned to that component $N_{dyn}$ by the dynamic decomposition methods.

In Fig.~\ref{fig:pureza_y_completitud} we show the results of calculating $C$ and $P$ for the disks and spheroids identified by the 6 dynamical decomposition methods for the 9 equilibrium galaxy models as a function of the stellar mass of the galaxy. The upper left panel shows $C$ of the disk while the upper right panel shows $C$ of the spheroid. The lower left panel shows the $P$ of the disk while the lower right panel shows the $P$ of the spheroid.

In all galaxies, the spheroids identified have $C$ roughly between 0.66 to 1.25 (Fig.~\ref{fig:pureza_y_completitud}, upper right panel). JHistogram and JEHistogram methods present the smallest variation of $C$ around 1, with differences lower than $\sim0.04$. GaussianMixture systematically identifies spheroids with $1.01\lesssim C\lesssim 1.25$, that is more massive, while KMeans systematically identifies spheroids with $0.91\lesssim C\lesssim 0.66$, that is less massive. Additionally, the values of $P$ are higher than 0.79 for spheroids in the case of the two less massive galaxies, and higher than 0.92 in the rest of the cases, irrespective of the method used (Fig.~\ref{fig:pureza_y_completitud}, lower right panel).

For disks, the differences of $C$ are reversed and exacerbated (Fig.~\ref{fig:pureza_y_completitud}, upper left panel). This is due to the difference in mass between the two stellar components, which can be seen in Table~\ref{tab:masas_y_f_sph}, that mis-assigning a particle in the less massive stellar component has a greater impact in $C$ than in the more massive component. GaussianMixture systematically identifies disks with $0.72\lesssim C\lesssim 0.89$, that is less massive than the originals, although it achieves $P$ values closest to 1 (Fig.~\ref{fig:pureza_y_completitud}, lower left panel). The $C$ for the KMeans disks is always greater than 1 and is the method that reaches the highest values of $C$. This is because the $f_\mathrm{dsk}$ for this method, in each galaxy model, is between 1 to 4 times the $f_\mathrm{dsk}$ original. Consequently, $P$ for these disks goes from 0.86 to 0.24, making this method the one with the worst performance. In addition, as with the spheroids, the JHistogram and JEHistogram methods have the smallest variations in $C$ for the disks with differences of less than 0.06. Of these last two, the JEHistogram method is the one with the highest $P$ values, since the differences of the values of $P$ and 1 are always smaller than 0.17, for all the simulated galaxy models analyzed in this work (see the red curve in the lower left panel, Fig.~\ref{fig:pureza_y_completitud}).

Therefore, taking into account all the above, we can conclude that the JEHistogram method achieves the best compromise between completeness and purity.

\subsection{Stellar components}

As the disks and spheroids exhibit different dynamics, the stellar particles of each one are located in different regions in the Lindblad diagram (see as an example Fig.~\ref{fig:limblad}). This normalized diagram corresponds to the parameter space where the dynamical decomposition takes place. The z-component of the specific angular momentum is normalized with the specific angular momentum of a circular orbit for a given energy value. Additionally, the specific total energy is normalized by the specific total energy of the most bounded particle. Then, as the dynamical decomposition methods used in this work separate the stellar components in different ways, this results in differences in the distribution of their particles in parameter space of the normalized Lindblad diagram. Therefore, this could be seen in the resulting spatial distribution of the components, as discussed below. We selected the model galaxy AGAMA-04 as an example of the differences that can arise when identifying disk and spheroid in a galaxy using different implementations of dynamical decomposition. Fig.~\ref{fig:resultado_en_el_espacio_de_parametros} shows the particle distributions of the original disk and spheroid (first column) of this galaxy, and those obtained using the 6 implementations to decompose them dynamically (six remaining columns). The top row corresponds to the disk component and the bottom corresponds to the spheroid component. Variations in this diagram are a consequence of the variations in completeness and purity achieved in each case. They will be translated into differences in the properties of the recovered components. Therefore, we analyzed the differences that appear and how they vary for this galaxy model.

In the case of the JThreshold method the assignment of stellar particles depends only on a cut-off circularity parameter $\epsilon_\mathrm{cut}$. Therefore adopting different values of this parameter may result in variations in the mass of the disk and spheroid, in their mass distribution and characteristic velocities. We have considered one of the most commonly used values of $\epsilon_\mathrm{cut} = 0.7$ \citep[e.g.][]{2014_Marinacci, 2014_Vogelsberger}. As can be seen in the second column in Fig.~\ref{fig:resultado_en_el_espacio_de_parametros}, we find that the bulk distribution of the particles in the normalized Lindblad diagram for the stellar components has a certain similarity to the originals. The main difference is the lack of the particles with $J_z/J_\mathrm{circ} < 0.7$ for the disk component and complementary the lack of the particles with $J_z/J_\mathrm{circ} \geq 0.7$ for the spheroid component.

The JHistogram method presents an improvement over the JThreshold, since it ensures an equal number of counter-rotating and co-rotating stellar particles. This is reflected in its ability to recover the total mass of the spheroid correctly and, consequently, the total mass of the disk (see Fig.~\ref{fig:fracciones_de_masa_de_galaxias_modelo}), regardless of the stellar mass of the model galaxy used. Although the co-rotating particles are selected so that the $J_z/J_\mathrm{circ}$ distribution of the spheroid is symmetric about $J_z/J_\mathrm{circ} = 0$, the selection is carried out randomly (see the almost simetric histogram labeled spheroid in the bottom inset panel of Fig.~2 of \cite{abadi_03}). This impacts the energy distribution of the stellar particles in the resulting components. The Fig.~\ref{fig:resultado_en_el_espacio_de_parametros} (in the third column) shows how, in the example galaxy, this results in the assignment to the spheroid of a group of stars of high circularity but with energies less bounded to those corresponding to the original spheroid.

\begin{figure*}
	\includegraphics[width=\textwidth]{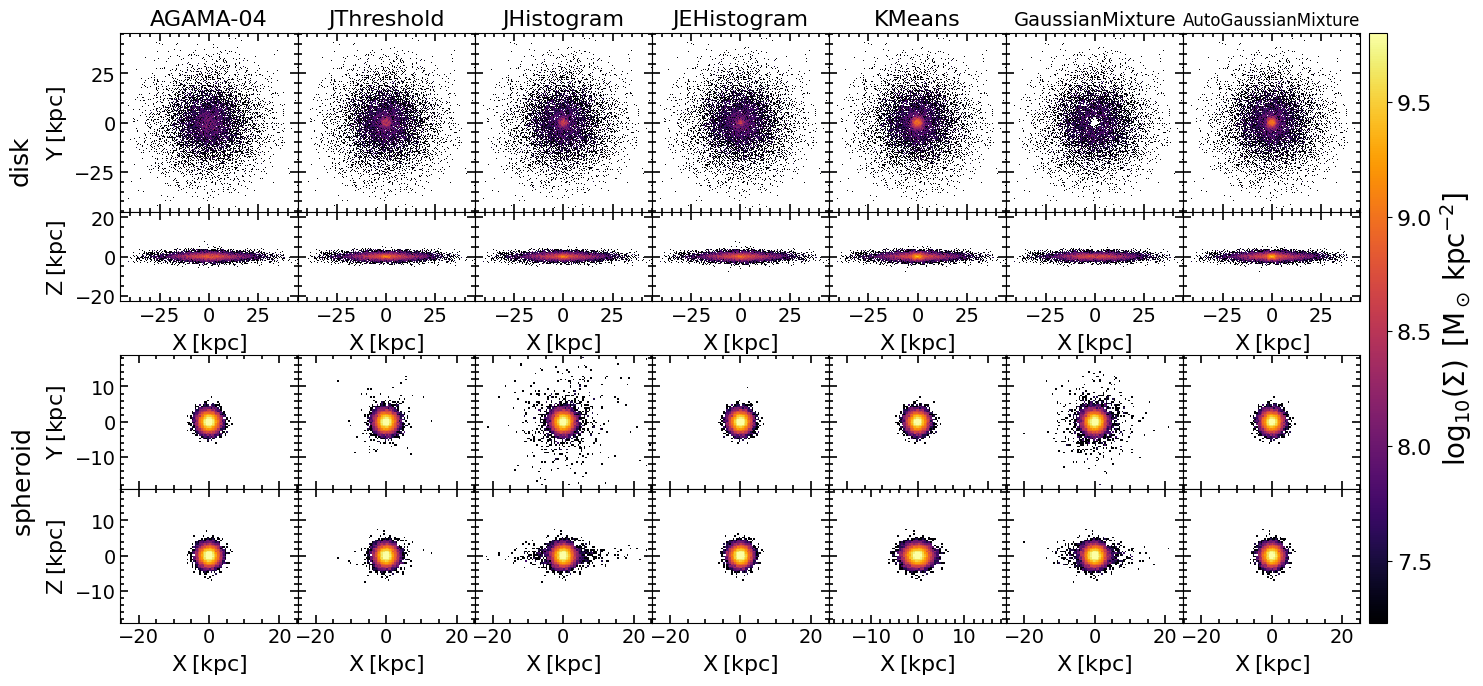}
    \caption{Face-on and edge-on view of the surface mass distribution of the disk (first row) and spheroid (second row) corresponding to the AGAMA-04 galaxy. The first column corresponds to the original stellar component distributions, while the rest corresponds to the components obtained by the 6 dynamical decomposition methods implemented. The color scale represents the value of the surface mass density.}
    \label{fig:distribucion_espacial_MW}
\end{figure*}

Compared to the JHistogram method, JEHistogram method considers the energy distribution of the counter-rotating particles to improve the assignment of the co-rotating particles to the two stellar components. This is mainly reflected in a better similarity in the original distribution of the particles in the Lindblad diagram especially that corresponding to the spheroidal component as shown in the case of the model galaxy in equilibrium, in the fourth column of the Fig.~\ref{fig:resultado_en_el_espacio_de_parametros}. Thus, taking into account the energy distribution improves the particle assignment. This, together with the fact that this method reproduces the mass fractions better (see Fig.~\ref{fig:fracciones_de_masa_de_galaxias_modelo}), indicates the ability of the method to recover the components correctly and, in consequence, to replicate the surface mass distribution and mass density profiles of the stellar components of galaxies, their characteristic velocity profiles, etc., as will be discussed later.

In the case of the KMeans, its performance depends on the number of particles in each component and their form in the space of parameters. This is because the method performs the separation by minimizing a criterion that measures how coherent the clusters are. Also assumes that the clusters are convex and isotropic, which is not always the case \citep[as an example see][]{Obreja_2016}. Therefore the lack of the ability to distinguish the stellar components of the galaxy reasonably increases with mass, due to the rising difference in the number of particles that form them. This results in less massive (more massive) spheroids (disks) than the original ones (see Fig.~\ref{fig:fracciones_de_masa_de_galaxias_modelo}) and also is reflected in the particle distribution of the resulting stellar components, as a lack (excess) of particles with high circularity in the spheroid (disk) (as we can see, for example, in the fifth column of Fig.~\ref{fig:resultado_en_el_espacio_de_parametros}).

\begin{figure}
\includegraphics[width=\columnwidth]{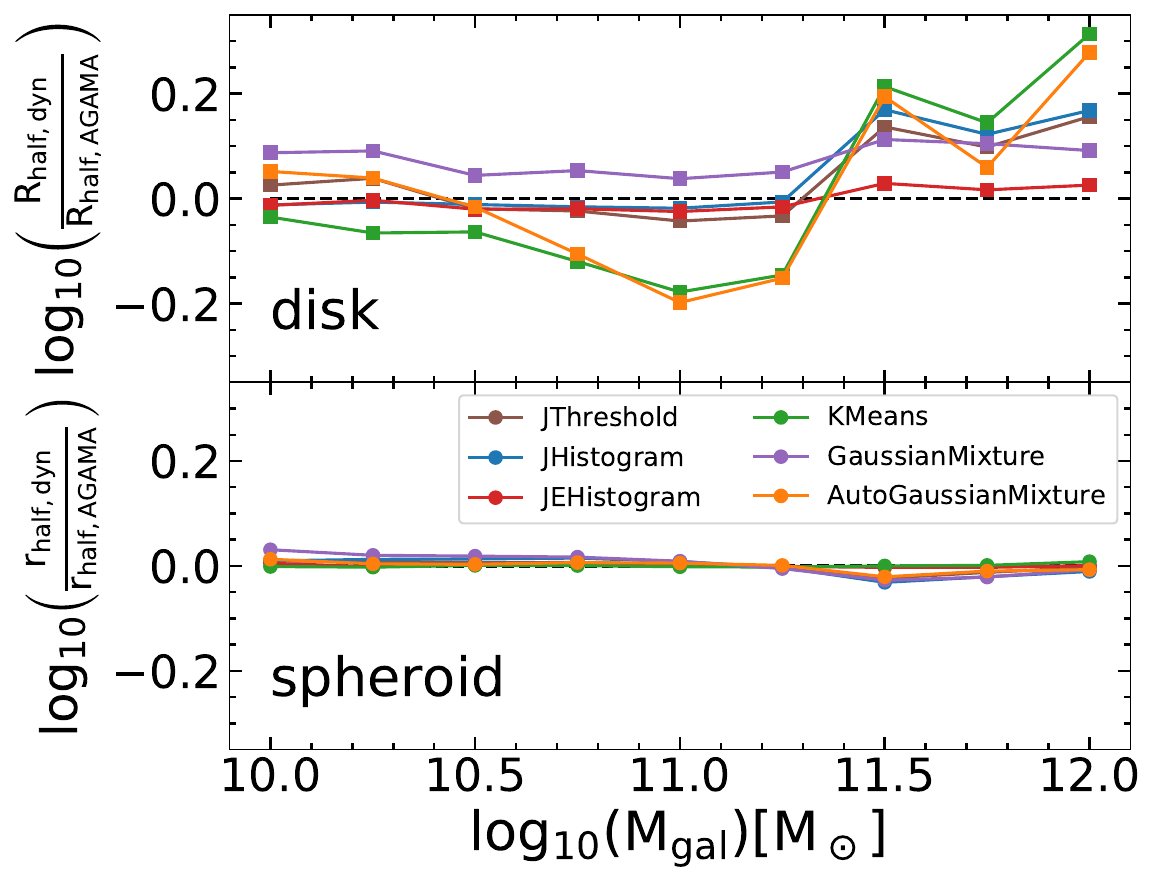}   
\caption{Comparison of the half mass radius obtained by the 6 dynamic decomposition methods and those of the original components of the 9 equilibrium galaxy models. Top panel: comparison of the half mass radius of the disks in cylindrical coordinates. Right panel: comparison of the half mass radius of the spheroids in spherical coordinates.}    
\label{fig:r_half_recuperado}
\end{figure}

The GaussianMixture method makes a probabilistic assignment of the particles because the method assumes that it is a mixture of Gaussians in the parameter space that produces the point distribution. It therefore assigns to each stellar particle the probability of belonging to one or the other stellar component, rather than assigning membership in a taxative way. Despite this, if the distribution of the stellar components in the Lindblad diagram does not have a Gaussian origin, the assignment of the particles to the disk or spheroid may not be the optimal. An example of this can be seen in the sixth column of Fig.~\ref{fig:resultado_en_el_espacio_de_parametros}, where the method seems to mis-assigning the most bounded particles with values of circularity $\sim1$. This impacts the mass of the component, resulting in spheroids (disks) that are more (less) massive than the originals (see Fig.~\ref{fig:fracciones_de_masa_de_galaxias_modelo} as an example), and also in his mass distribution.

\begin{figure*}
	\includegraphics[width=\textwidth]{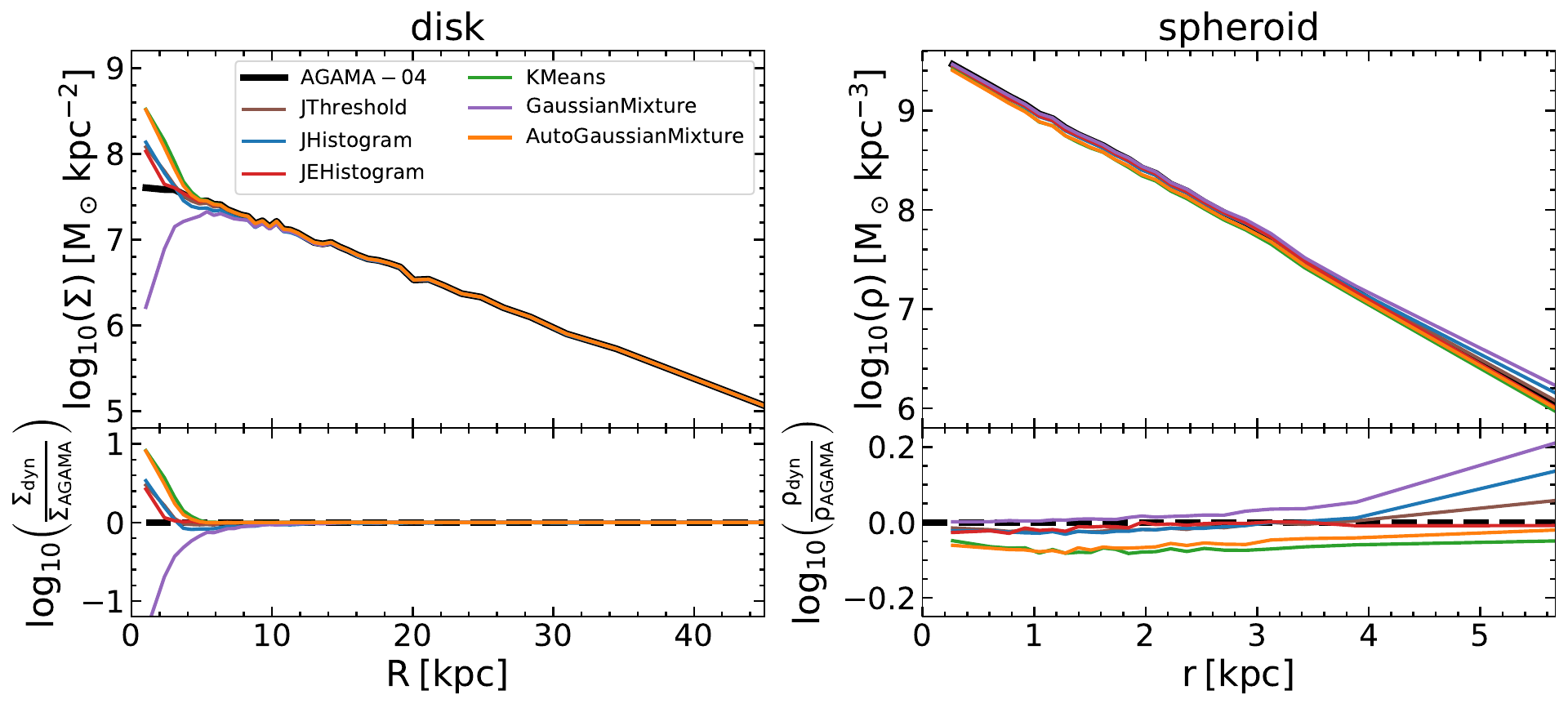}
    \caption{Surface mass density profiles of the original disk of the AGAMA-04 galaxy and those recovered by the 6 dynamical decomposition methods (upper left panel) and comparison between them (lower left panel), as a function of the cylindrical radius $R$. Mass density profiles of the original spheroid and those recovered by the different dynamical decomposition methods (upper right panel) and comparison between them (lower right panel) as a function of the 3D radius $r$. In all cases, we observe that the largest differences occur where both components coexist spatially.}
    \label{fig:perfiles_densidad}
\end{figure*}

Finally, let us analyze the resulting distributions for the components identified by the AutoGaussianMixture method. For this, we must take into account that this method identifies by default: halo, bulge, cold disk, and warm disk. Therefore we have decided to unify the halo and bulge in the spheroidal component, on the other hand, cold disk and warm disk in the discoidal component. This method, developed by \citet{du_19}, presents an improvement in recovering the stellar components concerning the previous method. This is because the particles are distributed in such a way that the use of a single Gaussian to characterize each component seems insufficient. An example of this behavior can be observed in the last column of Fig.~\ref{fig:resultado_en_el_espacio_de_parametros}, which is reflected in the fact that the particle distributions in the normalized Lindblad diagram of the recovered components are more similar to those of the original ones.

\subsection{Spatial mass distribution}\label{subsec:distribucion_espacial}

As an example of the result of the stellar components identified through the different methods of dynamic decomposition, we show in Fig.~\ref{fig:distribucion_espacial_MW} the spatial mass distribution of the disks and spheroids of one of the 9 equilibrium galaxy models included in Table~\ref{tab:masas_y_f_sph}. The top 2 rows show the spatial distribution of the particles identified as belonging to the disk component while the bottom 2 rows show those of the spheroid. The bins are colored according to the value of the amount of mass divided by the area of the bin, as indicated in the colorbar on the right hand side of the plot. From a visual inspection, we can see that the identified components look similar to each other and similar to the original ones. The differences are present where the components overlap spatially, which is in the center of the galaxy because here it becomes more difficult to distinguish to which component each stellar particle belongs. In the case of the disks, the edge-on views appear indistinguishable from each other, while in the face-on views, the disks appear to have an excess of mass density in the center, except for the case of the GaussianMixture method which has a hole in the center (see first and second rows of Fig.~\ref{fig:distribucion_espacial_MW}). In the case of the spheroids, also there do not seem to be many differences. The JThreshold, JHistogram, and GaussianMixture methods seem to show misclassified particles in the outer regions (see third and fourth row of Fig.~\ref{fig:distribucion_espacial_MW}).

To extend this analysis to a more quantitative one we decided to compare the stellar half-mass radius of the components obtained by all methods for each of the 9 model galaxies. In Fig.~\ref{fig:r_half_recuperado} we present the ratios between the half-mass radius of the component identified by dynamical decomposition and the half-mass radius of the corresponding original component. In the upper panel, the radii were measured in cylindrical coordinates with the disk viewed face-on, while in the lower panel, the radii were measured in spherical coordinates. First of all, regardless of the method used, the spheroids reproduce the radius at half mass with differences of less than 7.4\%. The JEHistogram method is the one that presents the smallest variations since in any case, for all simulated galaxy models, they do not exceed 1.2\%. If we now analyze the results for the disks, more notable variations appear. However, the JEHistogram method is still the one with the smallest differences in radius at half mass with respect to that of the original components, with values of less than 7.0\%.

\subsection{Component mass density profiles}\label{subsec:perfiles_de_masa}

\begin{figure}
	\includegraphics[width=\columnwidth]{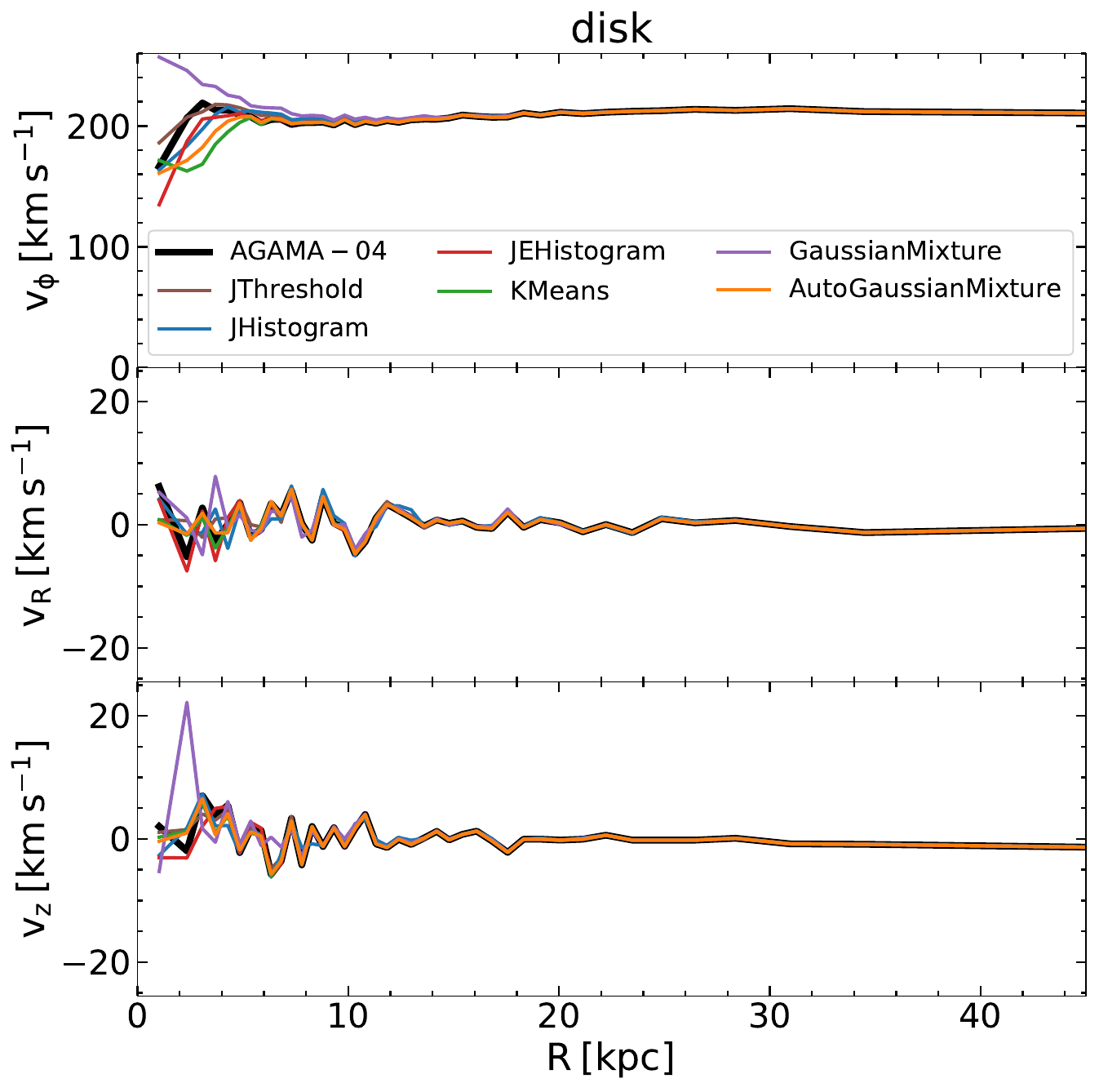}
    \caption{Comparison of velocity profiles as a function of cylindrical radius $R$ of the AGAMA-04 galaxy disk with those identified using the 6 different methods of dynamical decomposition. Top panel: comparison of rotational velocity $v_\phi$. Middle panel: comparison of the radial velocity profile $v_R$. Bottom panel: comparison of the vertical velocity profile $v_z$. In all cases, the solid black line corresponds to the velocity profile of the original stellar disk. These profiles are calculated with bins with an equal number of particles.}
    \label{fig:vrot-profile-agama}
\end{figure}

The mass density profiles of the disks are typically exponential \citep{Freeman_1970} while those of the spheroids are de Vaucouleurs profiles \citep{de_Vaucouleurs_1959}. Thus they serve to test the ability of different implementations of the dynamical decomposition to reproduce the mass density profiles of the stellar components of the galaxy. Therefore, we performed a comparison between the original mass density profiles and those 6 recovered by dynamic decomposition. For the disks, we calculate the radial surface mass density profiles in rings $\Sigma = m / \pi (R_\mathrm{out}^2 - R_\mathrm{in}^2)$  in a face-on projection (where $R_\mathrm{out}$ is the outer cylindrical radius of the ring, $R_\mathrm{in}$ is the inner cylindrical radius of the ring and m is the mass inside the ring) for the original stellar component $\Sigma_\mathrm{AGAMA}$ and the components identified with the dynamic decomposition $\Sigma_\mathrm{dyn}$. Also, we estimate the differences between the original disk surface mass density profile. For spheroids, we calculate the radial volumetric mass density profiles in spherical shells $\rho = 3m / 4\pi (r_\mathrm{out}^3 - r_\mathrm{in}^3)$, where $r_\mathrm{out}$ is the outer spherical radius of the shell, $r_\mathrm{in}$ is the inner spherical radius of the shell and m is the mass inside the shell. These profiles were calculated for the original stellar component $\rho_\mathrm{AGAMA}$ and the components identified with the dynamic decomposition $\rho_\mathrm{dyn}$. All profiles were calculated with bins of an equal number of particles. An example of this is shown in Fig.~\ref{fig:perfiles_densidad} which corresponds to the stellar components of the AGAMA-04 galaxy. The left panel shows the original profile with those of the obtained components (top) and the comparison (bottom) in the case of the disk. The profiles corresponding to the original spheroid and those obtained by the dynamic decomposition are on the top right panel while the comparison is on the bottom right. The part of the galaxy where it becomes difficult to identify to which component each stellar particle belongs corresponds to where the disk and the spheroid spatially overlap, as we can see in Fig.~\ref{fig:perfiles_densidad}. This refers mainly to the more central part where it is more difficult to distinguish to which component each stellar particle belongs and can be translated into an excess or deficit in the mass density profiles. For this galaxy, the mass profiles of spheroids identified dynamically show differences lower than 17.2\% in comparison with the original profile, except for the last bin where the JHistogram and GaussianMixture methods reach differences of 36.8\% and 62.5\% respectively. The JEHistogram method achieves the best performance, whereas differences are always lower than 6.5\%. This behavior is in agreement with the visual inspection in the two bottom rows in Fig.~\ref{fig:distribucion_espacial_MW}. The discs show differences between the original profile and the profiles of the components recovered by the methods are similar for $R > 4.28$ $\rm{kpc}$. In this range, the differences are less than 19.2\%, with the JEHistogram method achieving the best performances since differences are less than 3.5\%. The exception is the case of the GaussianMixture method which achieves a difference of 41.4\%. In the inner part of the disk, where the disk and spheroid coexist spatially, the impact of mis-assigning a particle is greater. Therefore in the first bin, the profiles obtained with the dynamic decomposition methods differ by approximately one order of magnitude from the original. Note that the imbalance between the mass of the spheroid and the disk must be taken into account to interpret how the misassign impacts the performance of the dynamical decomposition methods. This is because if the mass or number of particles of one of them is much greater than that of the other, the erroneous assignment of one of the stellar particles will have a greater impact on the less massive one.

\subsection{Profiles of characteristic velocity components}\label{subsec:velocidad_de_las_componentes}

\begin{figure}
	\includegraphics[width=\columnwidth]{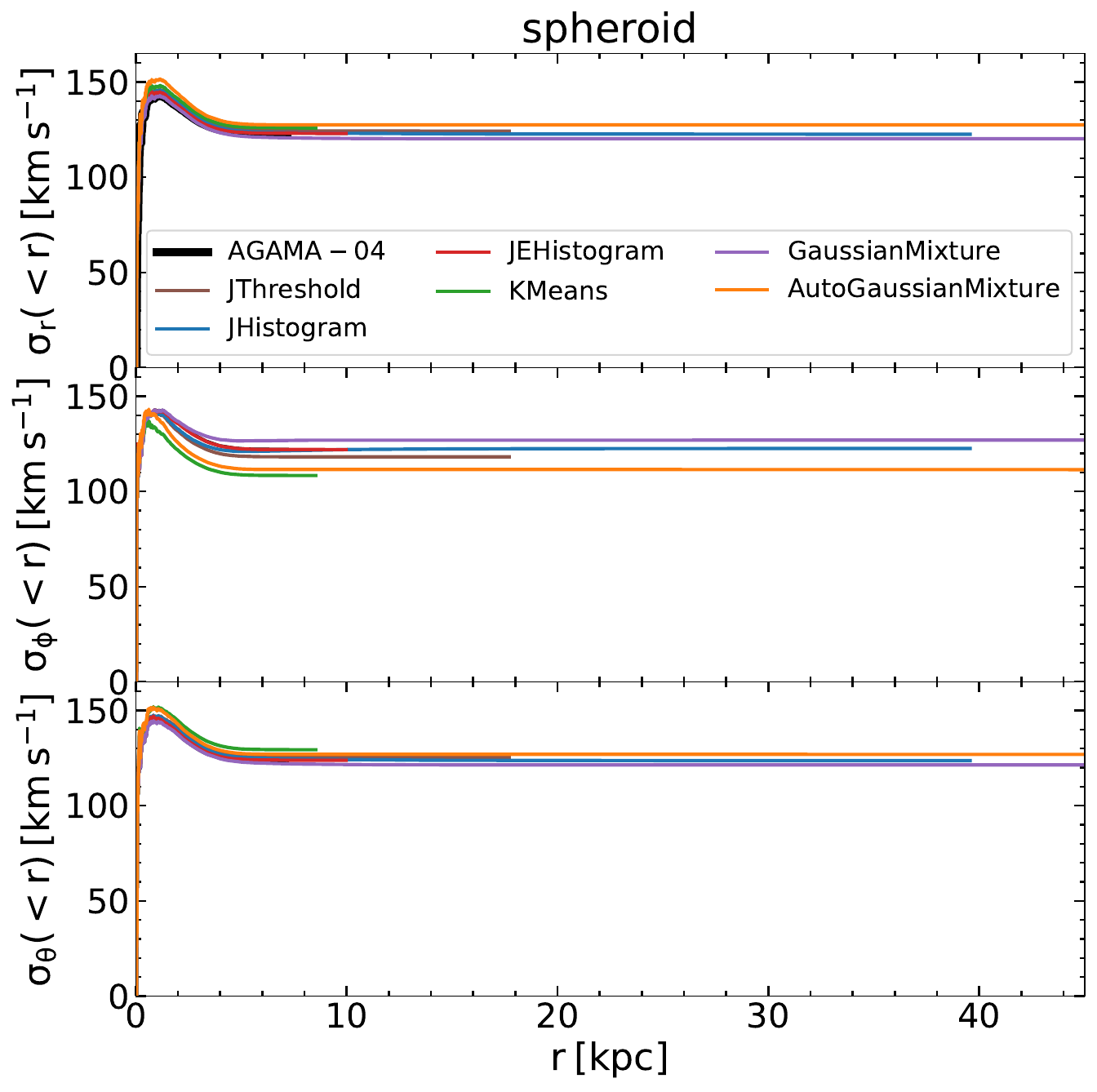}
    \caption{Integrated velocity dispersion profiles $\sigma_r(<r)$ (upper panel), $\sigma_\phi(<r)$ (center panel) and $\sigma_\theta(<r)$ (lower panel) directions for the spheroids as a function of spheroidal radius $r$ of the AGAMA-04 galaxy spheroid with those identified using the 6 different methods of dynamical decomposition. In all cases, the profiles corresponding to the original spheroid are in a black solid line.}
    \label{fig:sigma_spheroid-agama}
\end{figure}

Since the disks are supported by rotation while the spheroids are supported by velocity dispersion, the comparison of the recovered velocity profiles can provide an additional measure of the performance of the dynamic decomposition methods. To this end, we compute the velocity profiles in cylindrical coordinates $v_R$, $v_\theta$, $v_z$ for the discoidal components and the integrated velocity dispersion in spherical coordinates $\sigma_r(<r)$, $\sigma_\phi(<r)$, $\sigma_\theta(<r)$ for the spheroidal components, of one of the 9 models of galaxies in equilibrium as an example. On the one hand, the components of the velocities in cylindrical coordinates are given by
\begin{equation}
        v_R = (x\ v_x + y\ v_y) / R
\end{equation}
\begin{equation}
    v_\phi  = (x\ v_y - y\ v_x) / R
\end{equation}
where $R =\sqrt{x^2 + y^2}$, $x$, $y$, $z$ are the position of the particles in cartesian coordinates and $v_x$, $v_y$, $v_z$ are the velocities of the particles in cartesian coordinates. On the other hand, the integrated profiles of the components of the dispersion velocity in spherical coordinates are given by
\begin{equation}
        \sigma_j(<r) = \frac{1}{N(<r)} \sqrt{\sum_{i=1}^{N(<r)}(v_j(r_i) - \overline{v_j}(<r))^2}
\end{equation}
 where $j$ corresponds to one of the spherical coordinates, $N(<r)$ is the number of particles whose radius is less than r of the spheroidal component, $v_r=(x\ v_x + y\ v_y + z\ v_z) / r$, $v_\phi=(x\ v_y - y\ v_x) / \sqrt{x^2 + y^2}$, $v_\theta=(v_r\ z - v_z\ r) / \sqrt{x^2 + y^2}$, $v_j(r_i)$ is the velocity in direction $j$ of the i-th particle at radius $r_i$, and  $\overline{v_j}(<r)$ is the average velocity in direction $j$ of the particles inside the radius r. In Fig.~\ref{fig:vrot-profile-agama} we present the original and identified velocity profiles for the discoidal component of the galaxy model AGAMA-04. These profiles were calculated in bins with equal particle numbers. In addition, in Fig.~\ref{fig:sigma_spheroid-agama} we present the integrated velocity profiles as a function of the radius $r$ of the spheroidal components in the same galaxy model, also for the original and recovered components.

\begin{figure}
\includegraphics[width=\columnwidth]{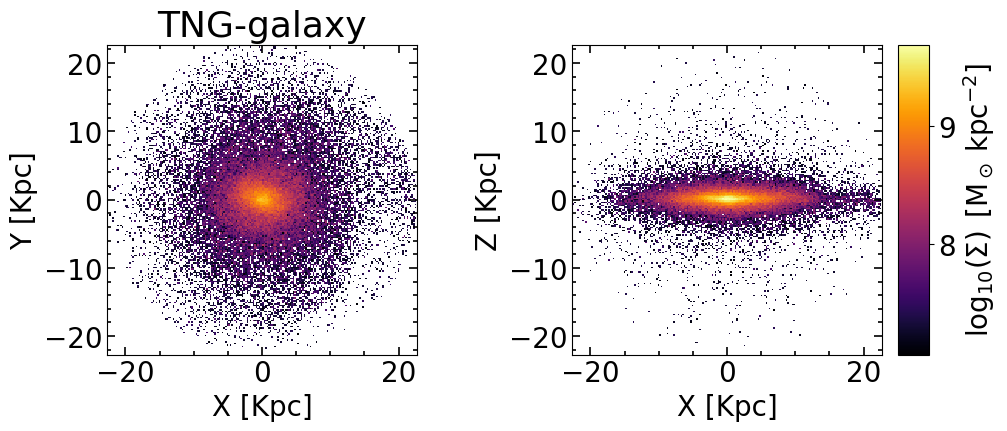}
\caption{Surface mass density of the stellar component for a galaxy from the IllustrisTNG simulations with a stellar mass and rotational velocity similar to those of the Milky Way (ID 486966). Left panel: face-on view of the galaxy. Right panel: edge-on view of the galaxy. In both cases, the presence of a disk and spheroid component is evident at first glance. The color scale corresponds to surface stellar mass density.}
\label{fig:tng_stars}
\end{figure}

We find that the discoidal components obtained using any of the employed methods exhibit $v_R$ profiles and $v_z$ profiles with values around zero, similar to the case of the original stellar disks as can be observed in the middle and bottom panels of the Fig.~\ref{fig:vrot-profile-agama}. These profiles generally assume values between $-7\ \mathrm{km\ s^{-1}}$ and $8\ \mathrm{km\ s^{-1}}$. Therefore, the reproduction of these profiles is robust regardless of the method employed for dynamic decomposition. In the case of rotational velocity profiles of the disks identified for this galaxy example (see top panel of Fig.~\ref{fig:vrot-profile-agama}), they exhibit differences compared to the original profiles whose relative errors are below 23.2\%, especially in the central region, where the disk and spheroid overlap in the galaxy. In the inner bin of the disk profile which corresponds to the GaussianMixture method, the value of the relative error raises 54.5\%. For all dynamical decomposition methods, these values fall to differences of 5\%. This shows the general robustness of the methods in obtaining discoidal components that manage to reproduce their characteristic velocity profiles.
Additionally, it confirms that is more difficult to recover the correct velocity profiles at the center.

Regarding spheroids obtained from dynamical decomposition for this galaxy model, we find that the integrated velocity dispersion profiles in the three directions are reproduced with differences of less than 11.2\% concerning the original profiles (see Fig.~\ref{fig:sigma_spheroid-agama}). In all cases, the different extensions of these velocity profiles reflect the position of the stellar particle most distant from the center of the stellar component. We note that the JEHistogram method is the one that most accurately reproduces these profiles because in the worst-case scenario, the differences are 1.5\%.

\begin{table}[t]
    \caption{Stellar mass and mass fraction of the disk and spheroid components, obtained by using the 6 dynamical decomposition methods implemented in the \gxchp{} package, of the simulated galaxy from the IllustrisTNG simulations (ID 486966).}
        \label{tab:masas_TNG}
    $$
        \begin{array}{m{0.30\linewidth}cccc}
            \hline
            Method  & M_{dsk} & M_{sph} & f_{sph} & f_{dsk}\\
              & [10^{10} M_\odot] & [10^{10} M_\odot] &  & \\
            \hline
            JThreshold          & 3.69 & 1.83 & 0.33 & 0.67 \\
            JHistogram          & 4.59 & 0.92 & 0.17 & 0.83 \\
            JEHistogram         & 4.60 & 0.92 & 0.17 & 0.83 \\
            KMeans              & 4.11 & 1.40 & 0.25 & 0.75 \\
            GaussianMixture     & 3.18 & 2.33 & 0.42 & 0.58 \\
            AutoGaussianMixture & 3.80 & 1.72 & 0.31 & 0.69 \\
            \hline
        \end{array}
    $$
\end{table}

\section{Analysis of stellar component properties in a galaxy from hydrodynamic simulation}
\label{sec:galaxychop}

\begin{figure*}[h]
\includegraphics[width=\textwidth]{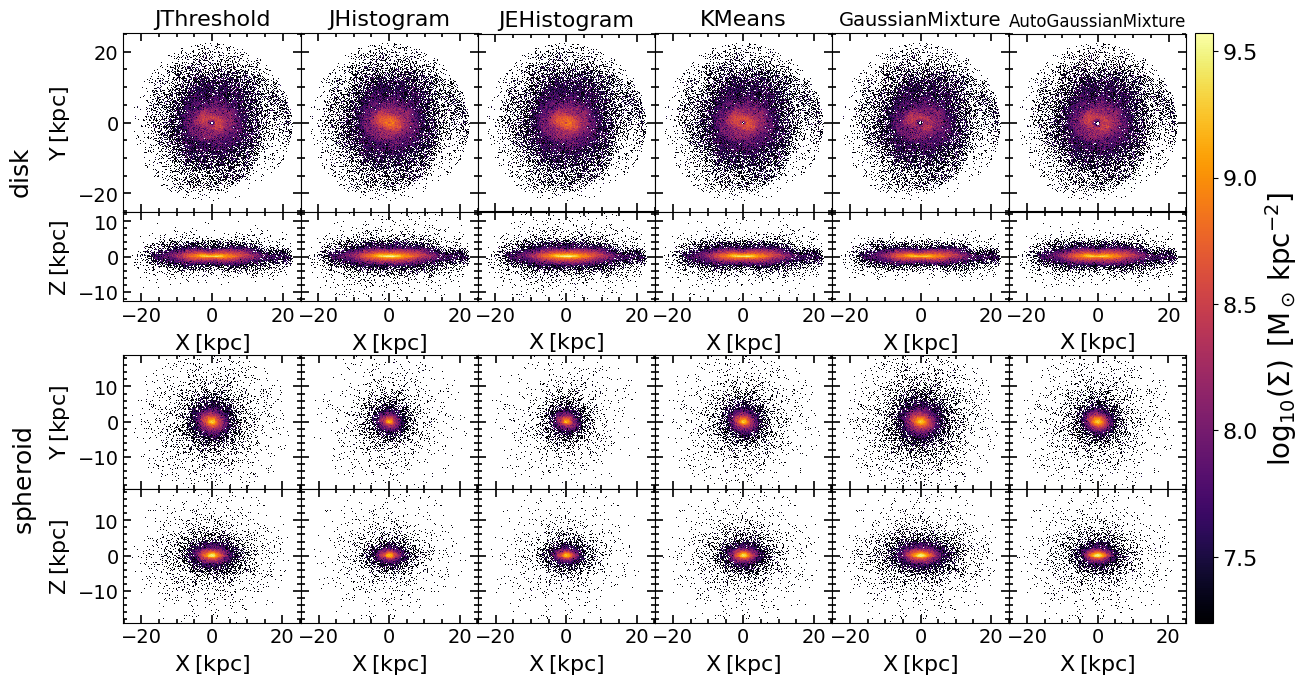}
\caption{Face-on and edge-on view of the surface mass distribution of the disk (first row) and spheroid (second row) corresponding to the IllustrisTNG galaxy. Each column corresponds to the stellar component distributions of the components obtained by the 6 dynamical decomposition methods implemented. The color scale represents the value of the surface mass density.}
\label{fig:densidad_de_masa_TNG}
\end{figure*}

In this section, we show an illustrative example of how the properties of the stellar disk and spheroid components can present variations in their properties when different dynamical decomposition methods are used in a galaxy belonging to a hydrodynamic simulation. For this purpose, we select a galaxy belonging to the IllustrisTNG hydrodynamic cosmological simulations \citep{Pillepich-2018-TNG, Naiman-2018-TNG, Nelson-2018-TNG, Marinacci-2018-TNG, Springel-2018-TNG}, which are state-of-the-art simulations. We have decided to choose a galaxy from the TNG100 run. The selected galaxy has a total stellar mass of $\rm{log_{10}(M_{gal}/M_\odot)=10.74}$, measured within $0.1 \times r_\mathrm{vir} = 22.7\ \mathrm{kpc}$, with $r_\mathrm{vir}$ being the radius of a sphere enclosing an average density 200 times the critical density of the universe, and a rotational velocity $v_\mathrm{rot}=219\ \mathrm{km\ s^{-1}}$ at half mass radius. Both properties were chosen to be similar to those of the Milky Way. The Fig.~\ref{fig:tng_stars} displays the distribution of the surface mass density of the galaxy face-on (left) and edge-on (right). It can be observed that the chosen galaxy, used for the subsequent comparison of the identified component properties using different methods of dynamical decomposition, exhibits a discoidal morphology. In addition to the disk component, a spheroidal component is observed in the form of an increased density in the central region of the galaxy. In this case we rotate the galaxy by aligning the z-axis with the spin of the galaxy calculated with the stellar particles that lie within $0.1 \times r_\mathrm{vir}$.

\begin{figure*}
\includegraphics[width=\textwidth]{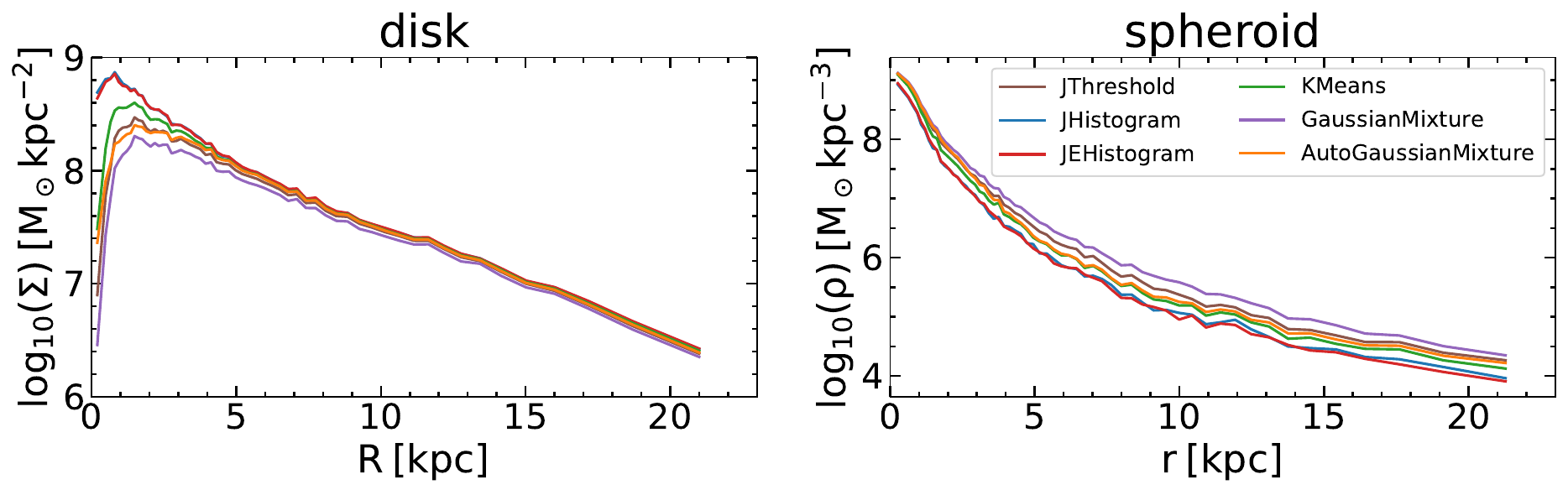}
\caption{Surface mass density profiles of the discoidal component of the IllustrisTNG galaxy, recovered by the 6 dynamical decomposition methods as a function of the cylindrical radius $R$ (left panel). Mass density profiles of the spheroidal component of the IllustrisTNG galaxy, recovered by the 6 dynamical decomposition methods as a function of the 3D radius $r$ (right panel). In all cases, we observe that the largest differences occur in the inner part of the disk.}
\label{fig:perfil_de_densidad_TNG}
\end{figure*}

We dynamically decomposed the selected galaxy using the 6 methods implemented in the \texttt{GalaxyChop} package. For the case of the JThreshold method, we use $\epsilon_\mathrm{cut} = 0.7$. In the KMeans and GaussianMixture methods, we use $n=2$ because we want to identify two stellar components: disk and spheroid. Finally, since the AutoGaussianMixture method identifies apriori 4 components we call the union of the cold disk and warm disk components a disk, while we call the union of the bulge and halo a spheroid. We compare properties such as stellar mass, surface mass distribution, mass density profiles, and characteristics velocity profiles of the resulting components. We also compare the mass density profiles with those obtained from classical photometric decomposition. Finally, we analyze some astrophysical properties such as their age and color.

\subsection{Mass fraction of the spheroid and mass distribution of the components}

In Table~\ref{tab:masas_TNG} we present the stellar masses and the mass fraction of the spheroids and disks identified, using the 6 dynamical decomposition methods implemented in the package. As we already saw in Sect.~\ref{subsec:fracc_de_masa} both disks and spheroids recovered by dynamical decomposition present differences in the amount of stellar mass assigned to them in each case. This results in the $f_\mathrm{sph}$ varying between $0.17 - 0.42$, with JHistogram and JEHistogram being the methods that recover the less massive spheroidal component.

The Fig.~\ref{fig:densidad_de_masa_TNG} shows the face-on and edge-on view of the surface mass distribution of the disks (top 2 rows) and spheroids (bottom 2 rows) obtained by dynamical decomposition to the IllustrisTNG galaxy. Each column corresponds to each of the 6 dynamical decomposition methods and the color scale represents the value of the surface mass density. Additionally, Fig.~\ref{fig:perfil_de_densidad_TNG} presents the mass density profiles of the stellar components of the IllustrisTNG galaxy recovered by the 6 dynamical decomposition methods. In the left panel are the surface mass density profiles of the discoidal component as a function of the cylindrical radius R, whereas in the right panel are the volumetric mass density profiles of the spheroidal component as a function of the 3D radius r.

\begin{table}[t]
    \caption{Disk half mass radius, disk half mass height, and spheroid half mass radius for the stellar components obtained with the dynamical decomposition methods implemented in the \gxchp{} package of the simulated galaxy from the IllustrisTNG simulations (ID 486966).}
        \label{tab:rz_half_TNG}
    $$
        \begin{array}{m{0.30\linewidth}cccc}
            \hline
            Method  & R_{half, dsk} & z_{half, dsk} & r_{half, sph}\\
                    & [kpc]         & [kpc]         & [kpc] \\
            \hline
            JThreshold          & 6.82 & 0.51 & 2.55 \\
            JHistogram          & 5.83 & 0.53 & 2.42 \\
            JEHistogram         & 5.88 & 0.53 & 2.28 \\
            KMeans              & 6.39 & 0.52 & 2.33 \\
            GaussianMixture     & 7.24 & 0.51 & 2.89 \\
            AutoGaussianMixture & 6.81 & 0.55 & 2.23 \\
            \hline
        \end{array}
    $$
\end{table}
     
\begin{figure}
\includegraphics[width=\columnwidth]{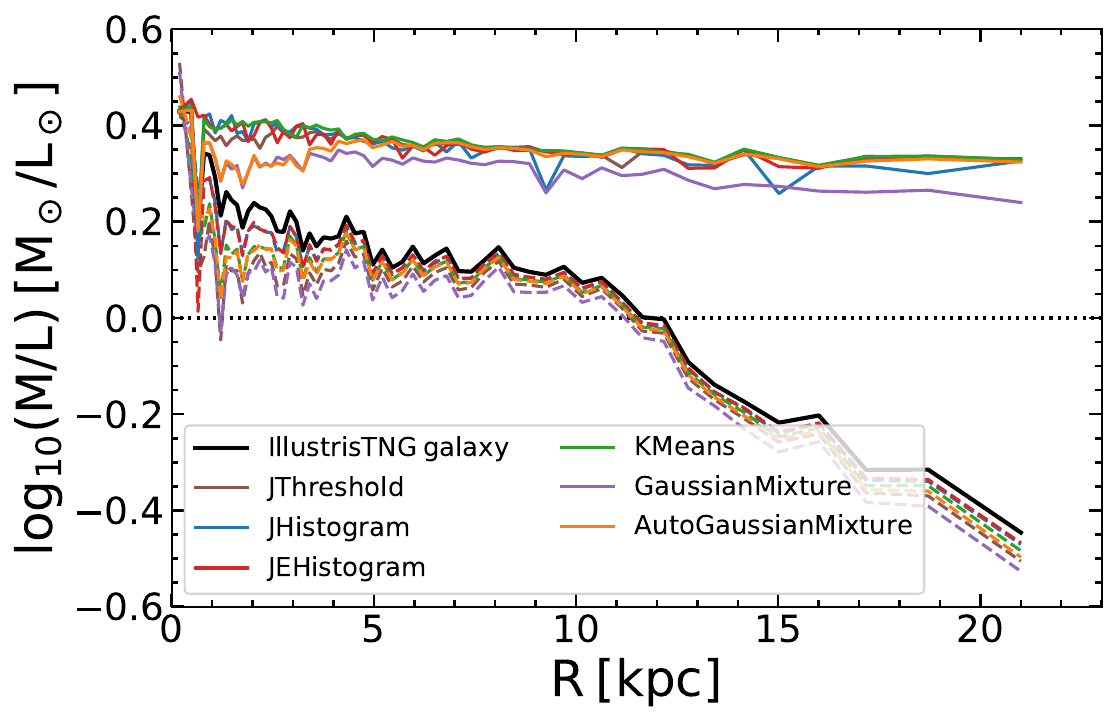}
\caption{Ratio between the radial surface density profile of the mass and the luminosity of the galaxy, calculated with bins with an equal number of particles, as a function of cylindrical radius $R$ of the IllustrisTNG galaxy. The black solid line is the mass-luminosity relation considering all the stellar particles. The solid lines correspond to the mass-luminosity relation for spheroids identified with the 6 methods of dynamic decomposition, while the dashed line corresponds to the discs. The dotted line is the one-to-one ratio of these quantities.}
\label{fig:M_vs_L}
\end{figure}

From a visual inspection of the surface mass density of the disks, we can see that the edge-on distribution seems to be very similar regardless of the dynamic decomposition method used. In the case of the face-on view, some differences appear as gaps in the central region, except for the JHistogram and JEHistogram methods. The presence of this drop in density in the center of the disks is also observed in the surface mass density profile (see left panel of Fig.~\ref{fig:perfil_de_densidad_TNG}). This behavior, which we already saw in the face-on view of the disk identified by the GaussianMixture method (Fig.~\ref{fig:distribucion_espacial_MW}) and in his corresponding surface mass density profile (Fig.~\ref{fig:perfiles_densidad}), is due to the misclassification that the methods have when assigning a particle from the galaxy central region to the disk or spheroid. If we analyze visually the surface mass densities of the spheroids, we find a roughly similar distribution between the 6 as can be seen in the last two rows of Fig.~\ref{fig:distribucion_espacial_MW}. This similarity is reflected in the volumetric density profiles of these stellar components since the differences observed between the profiles do not exceed 0.63 in the logarithm in any case.

Finally, to complete the analysis in a quantitative way, we measured the radii at half stellar mass in cylindrical coordinates and the height at half mass of the disks, and the radii at half stellar mass in spherical coordinates of the spheroids. The values obtained are presented in Table~\ref{tab:rz_half_TNG}. We find that the radii for the disks differ by less than $1.41\ \mathrm{kpc}$. These differences are due to different density profile decreases in the central region of the disks, as we can see in the left panel in Fig.~\ref{fig:perfiles_densidad}, since the deeper the profile drop, the larger the radius at half mass the component will have. In addition, the height at half mass of the disks differs by less than $0.04\ \mathrm{kpc}$ when we compare the values obtained by the 6 dynamic decomposition methods. For spheroidal components, the radii at half mass differ by at most $0.66\ \mathrm{kpc}$. This further reinforces the idea that the results obtained are robust regardless of the dynamical decomposition method used to identify the stellar components, except for the half-mass radius of the discoidal component.

\subsection{Photometric versus dynamical decomposition}

\begin{figure}
\includegraphics[width=\columnwidth]{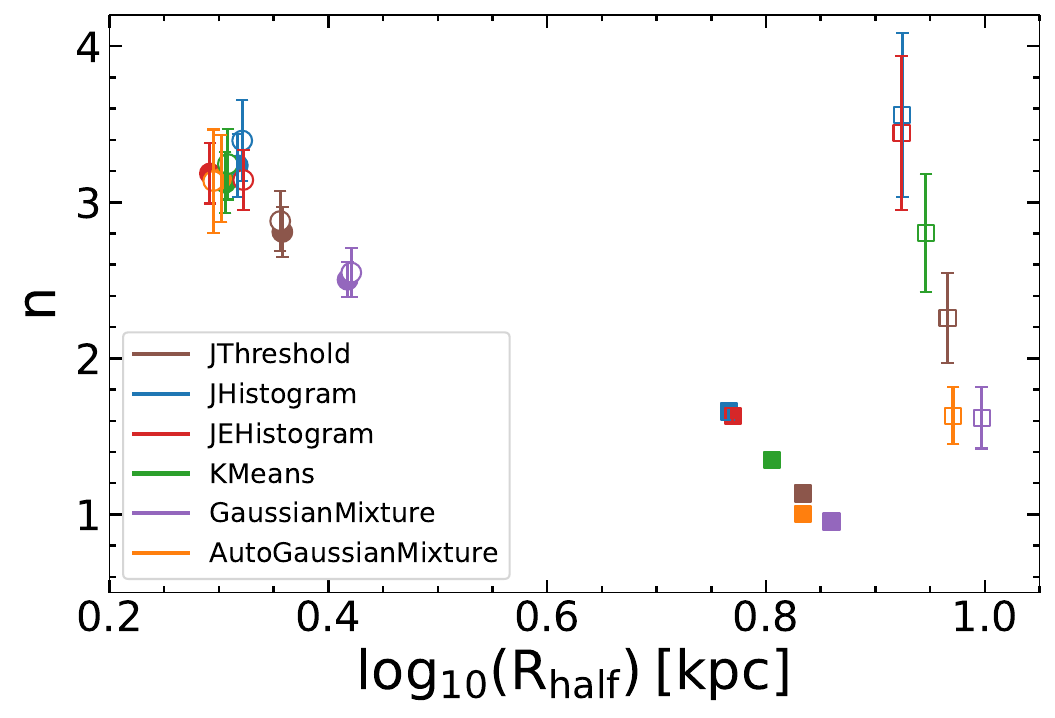}
\caption{Sérsic index versus cylindrical half mass radius $R_\mathrm{half}$ for disks and spheroidal half mass radius $r_\mathrm{half}$ for spheroids. The squares represent values obtained for disks and circles for spheroids. On the other hand, the filled symbols represent quantities obtained from the mass-weighted density profiles while the empty symbols correspond to quantities obtained from the light-weighted density profiles. Each color represents the stellar components obtained with the 6 dynamic decomposition methods implemented. Note that the error bars for the filled squares are smaller than the size of the symbols.}
\label{fig:n_vs_Re}
\end{figure}

Traditionally, to distinguish between the different stellar components of a galaxy, the distribution of their light is usually taken into account. This is done by fitting the radial profile of the total surface brightness through the sum of the radial profiles of the stellar components. The radial surface brightness profile of the spheroidal component follows a de Vaucouleurs profile \citep{de_Vaucouleurs_1959}, while on the other hand, the disk light distribution is given by an exponential type profile \citep{Freeman_1970}. Both profiles are generalized in the Sérsic profile \citep{sersic_1963}. However, only taking into account the spatial distribution of the stars and ignoring their dynamics makes it much more difficult to elucidate to which component a star in the central region belongs due to the overlapping of the stars. Additional factors that contribute to the complexity of photometric decomposition include the alignment of galaxies to the plane of the sky. This is because a single galaxy, when oriented differently, exhibits a distinct surface brightness distribution. In addition, we have the phenomenon of seeing, causing the blurring of surface brightness distribution, as well as extinction, which can notably fade the surface brightness within specific galaxy regions, further amplifying this complexity. It may also happen that the light coming from different regions of the galaxy does not reflect the amount of mass found in that region.

To explore the differences obtained from analyzing the light and mass distribution of the stellar components, we compared the mass and light density profiles for the 6 dynamical decomposition methods implemented in the \gxchp{} package. For this purpose, we fit a Sérsic profile to each luminosity-weighted and mass-weighted surface density profile, obtaining the values of the $n$ index in each case. All profiles were calculated as a function of the cylindrical radius R, from the face-on projection of the components and using bins of an equal number of particles. Moreover, we calculated the cylindrical radii enclosing half the mass and half the light of disks and spheroids. The results are shown in the Fig.~\ref{fig:n_vs_Re}.

The first thing we observe in the case of spheroids is that, given one of the dynamical decomposition methods, the values of $n$ and $R_\mathrm{half}$ are very similar regardless of whether light or mass is considered. This is because the mass-luminosity relationship of the spheroids is constant throughout, as can be seen in Fig.~\ref{fig:M_vs_L}, so the light distribution reflects the mass distribution. The values of $R_\mathrm{half}$ are very similar and independent of which dynamical decomposition method was used to identify the component. They range from $1.96\ \mathrm{kpc}$ to $2.63\ \mathrm{kpc}$. On the other hand, the $n$ values are around $n~\sim3$. 

\begin{figure}
	\includegraphics[width=\columnwidth]{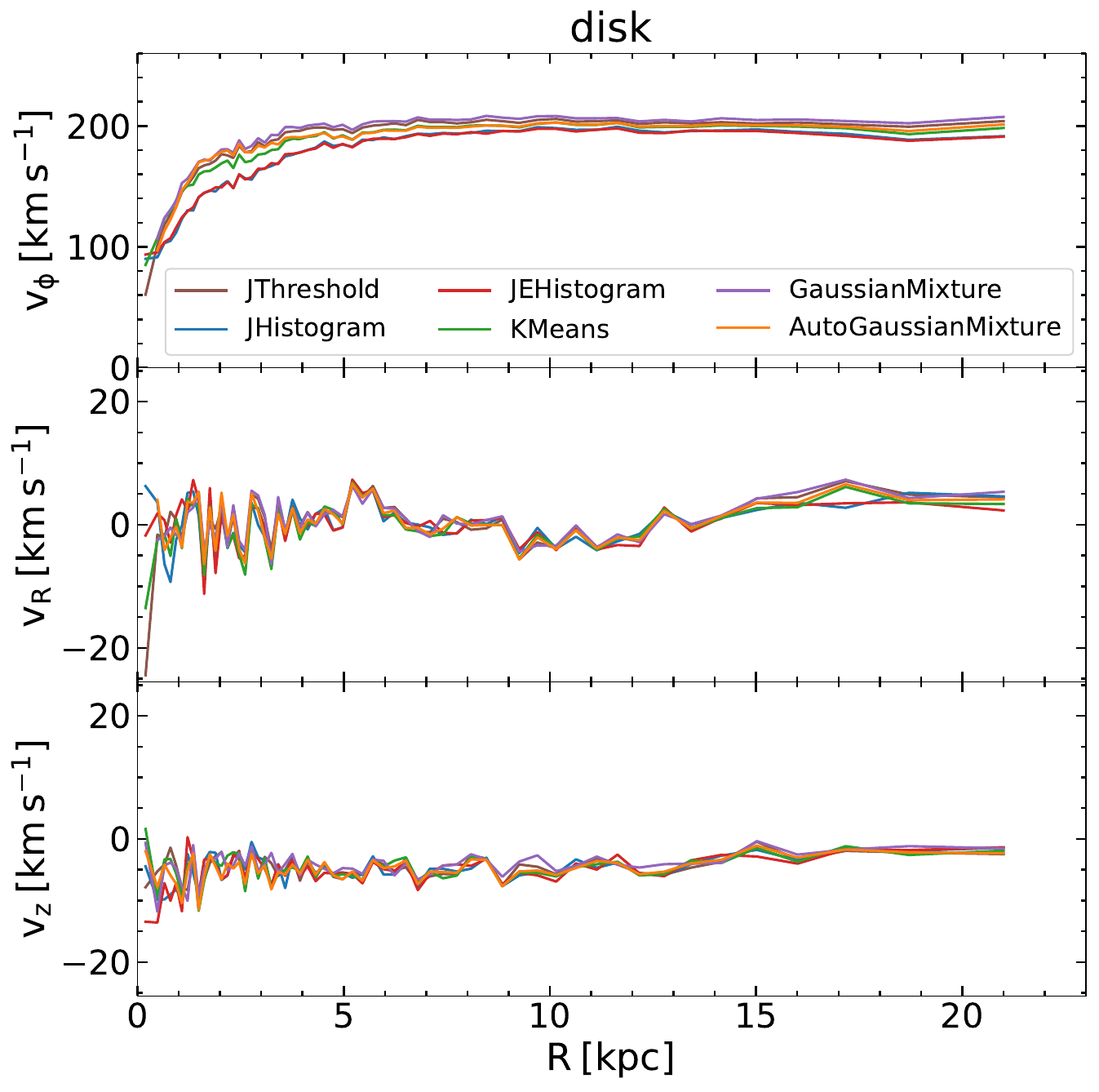}
    \caption{Comparison of velocity profiles identified using the 6 different methods of dynamical decomposition as a function of cylindrical radius $R$ of the IllustrisTNG galaxy disk. Top panel: comparison of rotational velocity $v_\phi$, the differences are more evident in the internal part of the disk. Middle panel: comparison of the radial velocity profile $v_R$. Bottom panel: comparison of the vertical velocity profile $v_z$. These two profiles adopt values around zero regardless of the method used.}
    \label{fig:perfil_comp_disk}
\end{figure}

\begin{figure}
	\includegraphics[width=\columnwidth]{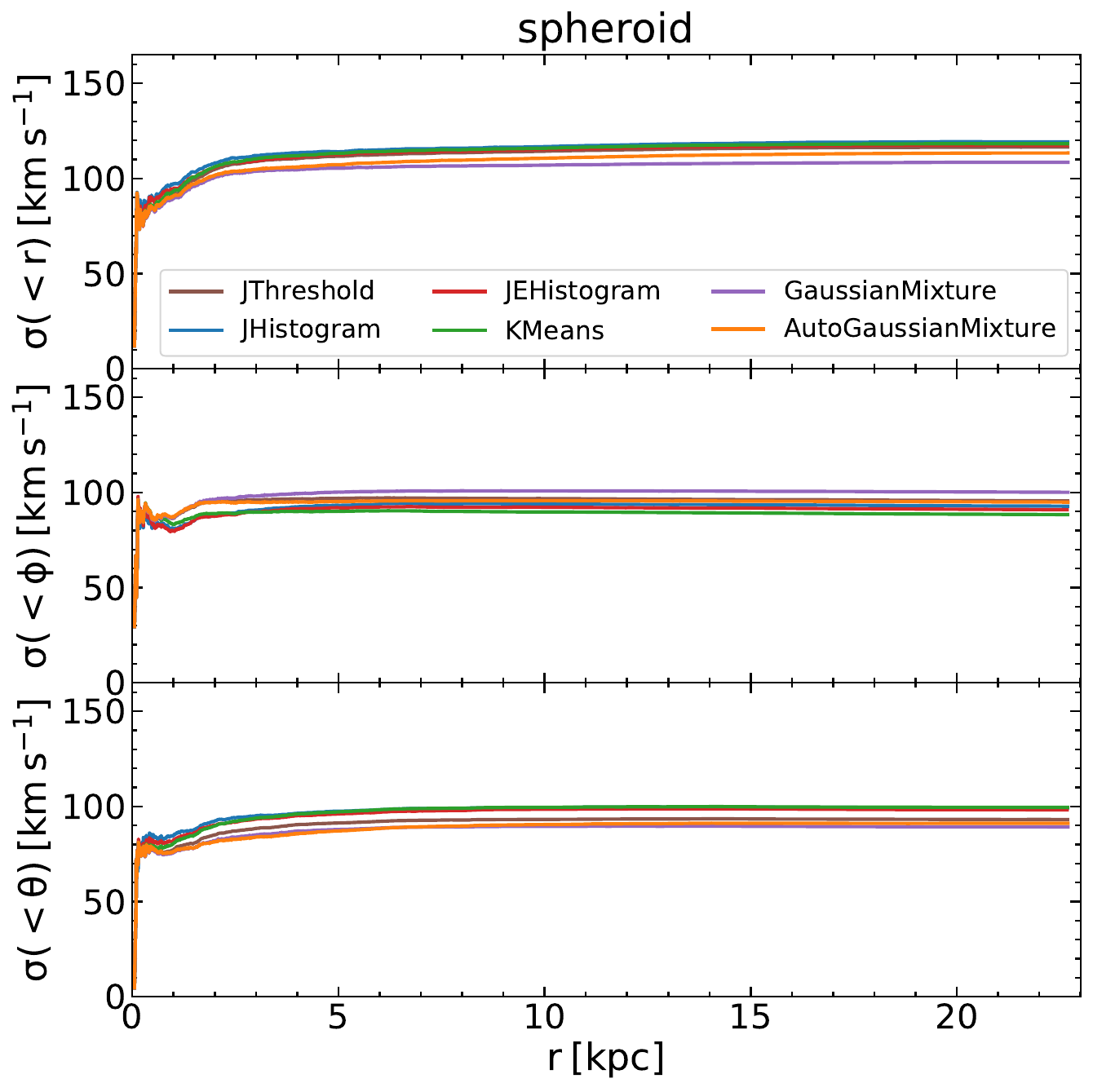}
    \caption{Integrated velocity dispersion profile $\sigma_r(<r)$ (upper panel), $\sigma_\phi(<r)$ (center panel), and $\sigma_\theta(<r)$ (lower panel) for the spheroids identified using the 6 different methods of dynamical decomposition, as a function of the 3D radius $r$ of the IllustrisTNG galaxy. In all cases, the profiles corresponding to the original spheroid are in a black solid line.}
    \label{fig:sigma_spheroid}
\end{figure}

In contrast, when we perform this analysis for the disks, we find that the values of $n$ and $R_\mathrm{half}$ present wide differences whether the light or mass distribution is considered. We find that the range of $R_\mathrm{half}$ for the light profiles is between $8.39\ \mathrm{kpc}$ and $9.93\ \mathrm{kpc}$, while for the mass profiles, the range of values goes between $5.83\ \mathrm{kpc}$ and $7.24\ \mathrm{kpc}$, depending on the dynamical decomposition method used (Fig.~\ref{fig:n_vs_Re}). This indicates that the light profiles are less concentrated than the mass profiles. The difference is because the mass-to-light ratio of the discoidal components varies with radius (see Fig.~\ref{fig:M_vs_L}). In addition, we find a large scatter in the values of the Sérsic indices obtained using the light distribution ($n \sim 1.62 - 3.56$), compared to those obtained from the mass distribution ($n \sim 0.95 - 1.66$). As already mentioned, these discrepancies are because the light and mass distributions of the galaxy vary along the projected radius (see Fig.~\ref{fig:M_vs_L}). In the outer regions where the discoidal component predominates, the light contribution increases compared to the mass causing the $M/L$ ratio of the galaxy to drop from $\sim 11\ \mathrm{kpc}$. Another thing we can notice, analyzing the mass-luminosity profile of the whole galaxy, is that the largest contribution in light, like the mass, is due to the disk component. This is due to the similarity between the mass-luminosity profiles of the entire galaxy and the disk, regardless of the method used to identify the stellar components of the galaxy.

These results indicate that the components obtained from the typical photometric decomposition are very different from those obtained from the dynamic decomposition, mainly because the light distribution of the stars does not always reflect the mass distribution of the stars. In the simulations, this behavior seems to be independent of the dynamical decomposition process by which the identification of the stellar components was carried out.

\subsection{Characteristic velocity profiles}

We compare the velocity profiles in cylindrical ($v_R, v_\phi, v_z$) and the integrated velocity dispersion in spherical coordinates ($\sigma_r(<r), \sigma_\theta(<r), \sigma_\phi(<r)$) for the disk and the spheroid respectively. Both profiles were obtained similarly to those presented in Sect.~\ref{subsec:velocidad_de_las_componentes}. Figure~\ref{fig:perfil_comp_disk} shows $v_\phi$ profile at the top panel, $v_R$ profile at the center panel, and $v_z$ profile at the bottom panel. There it can be seen that regardless of the method $v_R$ and $v_z$ assume values around zero with variations that do not generally exceed $10\ \mathrm{km\ s^{-1}}$ in absolute value. In the case of the $v_\phi$ (top panel of Fig.~\ref{fig:perfil_comp_disk}) the profiles are similar, regardless of the method used. For $R \lesssim 5\ \mathrm{kpc}$ the differences are less than 20\% (which represents $\sim 30\ \mathrm{km\ s^{-1}}$), while for $R \gtrsim 5\ \mathrm{kpc}$ the differences are reduced to 9\% (which represents $\sim 20\ \mathrm{km\ s^{-1}}$). This is the same behavior that we had observed in the model galaxy (Fig\ref{fig:vrot-profile-agama}, in Sect.~\ref{subsec:velocidad_de_las_componentes}), and it is a consequence of the fact that in the central part of the galaxy, it is difficult to assign the stellar particles. The integrated velocity dispersion profiles are shown in Fig.~\ref{fig:sigma_spheroid}. $\sigma_r(<r)$ are shown in the upper panel, the $\sigma_\theta(<r)$ are show in the middle panel and the $\sigma_\phi(<r)$ are show in the bottom panel. As in the case of the spheroid of the equilibrium galaxy model, all methods identify spheroids with integrated velocity dispersion profiles similar to each other. Regardless of direction, profiles have variations of up to 12\%. For $r \lesssim 2\ \mathrm{kpc}$ a small peak appears, which corresponds to the difficulty of correctly assigning the particles of the innermost region of the components.

It follows that the differences in the identified components with different implementations of dynamical decomposition methods would have little impact on their characteristic velocity profiles. This impact depends on how the methods assign the particles to the inner region of the galaxy, where both stellar components coexist.

\begin{figure}
	\includegraphics[width=\columnwidth]{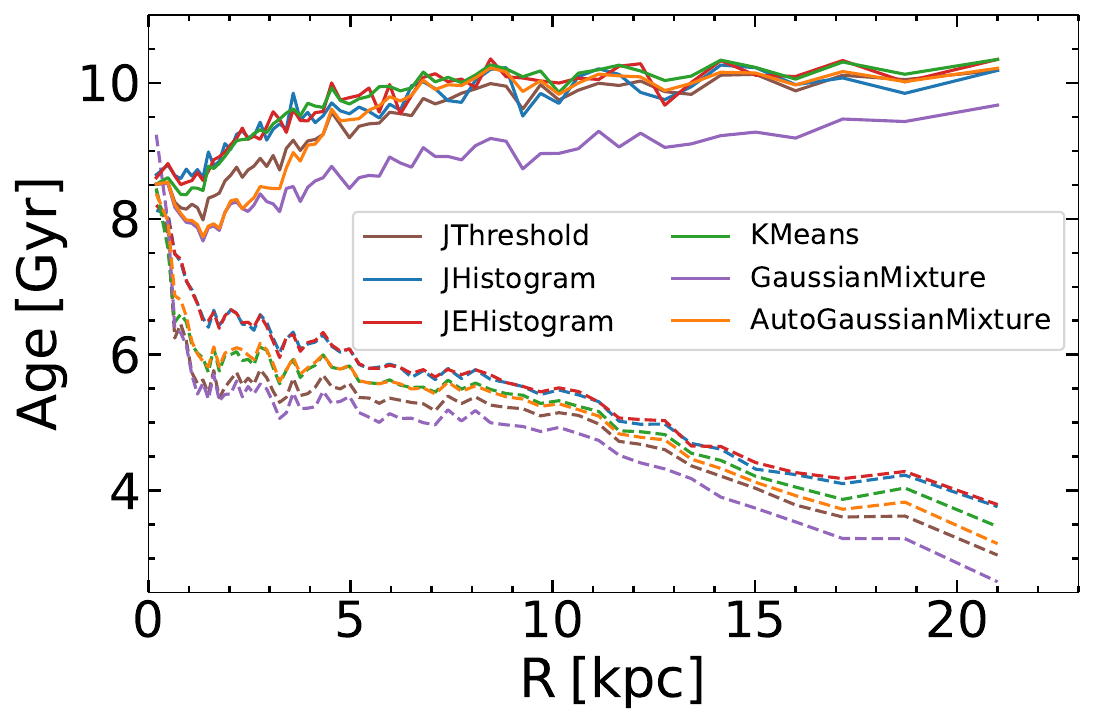}
    \caption{Comparison of the radial age profiles of the stellar components of the galaxy from IllustrisTNG simulations for the 6 dynamical decomposition methods as a function of cylindrical radius $R$. The dashed lines are the profiles corresponding to the disk, while the solid lines are the profiles corresponding to the spheroid.}
    \label{fig:age_TNG}
\end{figure}

\subsection{Age and color profiles of stellar components}

\begin{figure}
	\includegraphics[width=\columnwidth]{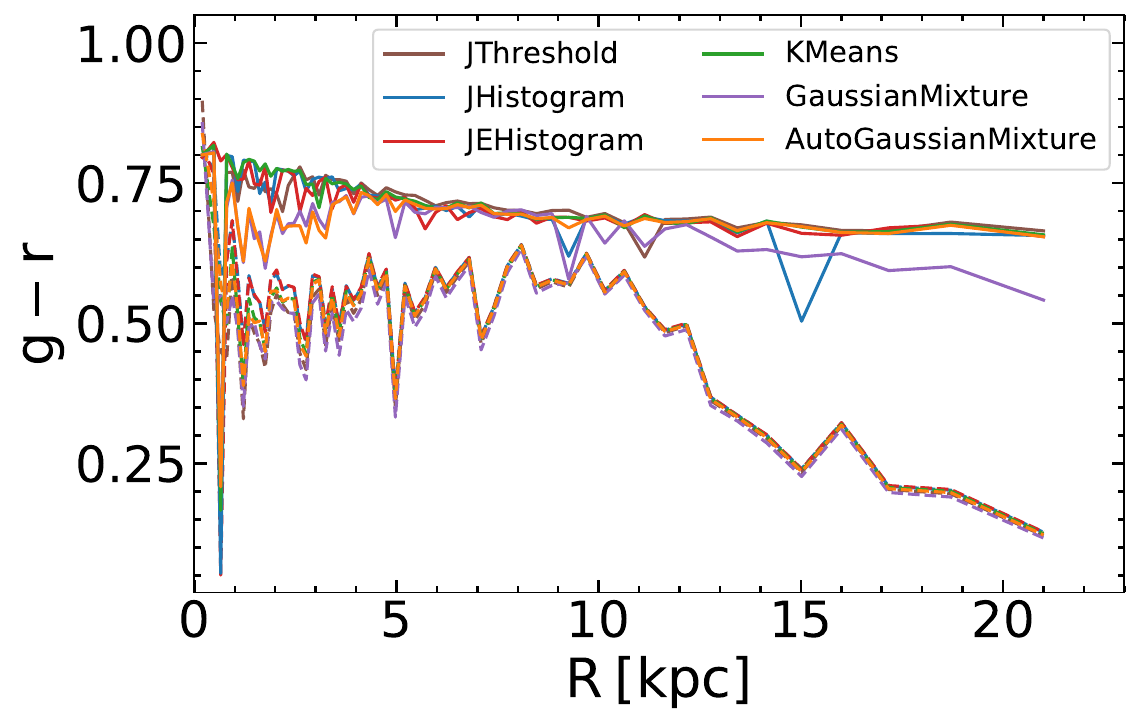}
    \caption{Comparison of the radial color g-r profiles of the stellar components of the galaxy from IllustrisTNG simulations for the 6 dynamical decomposition methods as a function of cylindrical radius $R$. The dashed lines are the profiles corresponding to the disk, while the solid lines are the profiles corresponding to the spheroid.}
    \label{fig:g_r_TNG}
\end{figure}

Age and color are two fundamental intrinsic properties that provide information about galaxy evolutionary processes. Here we will analyze the relationship between the identified stellar components' dynamics and their astrophysical properties. For this purpose, we calculated the age and g-r color profiles of the disks and spheroids identified by the 6 dynamic decomposition methods implemented. These profiles were calculated in bins of cylindrical radius R of equal number of particles, in the face-on projection. The Fig.~\ref{fig:age_TNG} shows the comparison of the age profiles corresponding to the disk in dashed lines and that of the spheroids in solid lines. The g-r color profiles for the discs in dashed lines and for the spheroids in continuous lines are presented in Fig.~\ref{fig:g_r_TNG}.

We find that all the profiles show a similar trend regardless of the method used. The scatter in the profiles, between the methods is $< 0.5\ \mathrm{Gyr}$, except for the GaussianMixture method in the spheroid profile. This could be because the method is assigning an excess of particles to the spheroid that the rest of the methods assign to the disk (as can be seen from the drop in the disk mass density profile in the left panel of Fig.~\ref{fig:densidad_de_masa_TNG}). The result is a bluer spheroid and a bluer disk compared to the rest. For $R \gtrsim 5\ \mathrm{kpc}$ the components become indistinguishable, indicating that the methods give robust results to each other, while for $R \lesssim 5\ \mathrm{kpc}$ the differences are a consequence of the possible error in the assignment of the particles to one stellar component or the other because, in the central part of the galaxy, both components coexist spatially.

In the case of the disks, they all show a negative radial gradient, from the center to the outside of the disk. This gradient changes its slope at $R \sim 11\ \mathrm{kpc}$ and becomes more pronounced for $R \gtrsim 11\ \mathrm{kpc}$. This gradient in age is responsible for the change in the $M/L$ ratio which has a change in slope on approximately the same scale (see Fig.~\ref{fig:M_vs_L}). We also analyzed the vertical age profiles of the disks and found that they decrease as we approach the plane of the galaxy, suggesting the presence of a thin disk and a thick disk.

On the other hand, the spheroids have a positive radial gradient (see Fig.~\ref{fig:age_TNG}). These profiles are less pronounced than disk profiles. These profiles also change his slopes, becoming flatter at $R \gtrsim 5\ \mathrm{kpc}$, as a result of the difficulty that the methods have for assigning the stellar particles and that it affects the properties of the less massive stellar component. Another possible explanation for this phenomenon is that the spheroids are the result of the superposition of the bulge and the halo. This results in a slightly younger population in the center and an older population extending to larger radii. As for the vertical age profiles of the spheroids, there is only a variation from younger to older stars as we move away from the plane of the galaxy within the first $5\ \mathrm{kpc}$ of radius.

In the case of g-r color profiles (see Fig.~\ref{fig:g_r_TNG}), as is the case for all the properties discussed so far, we find that regardless of the dynamic decomposition method applied, the profiles are similar to each other for both the disk and the spheroid. The spheroid profiles are redder than the disk profiles, independently of the radius $R$. Also, the spheroid profiles have an approximately constant value of $g-r \sim 0.74$, with a scatter of $\sim 0.3$ in the inner part and $\sim 0.03$ in the outer part except for the GaussianMixture, JHistogram, and JThreshold methods.

As in the case of the age profile, the $g-r$ color profiles recovered with all methods show two different behaviors for $R \lesssim 11\ \mathrm{kpc}$ and $R \gtrsim 11\ \mathrm{kpc}$. On the one hand, for $R \lesssim 11\ \mathrm{kpc}$, the profile is approximately constant value of $g-r = 0.55$ and a scatter of $0.05$. On the other hand, for $R \gtrsim 11\ \mathrm{kpc}$, it becomes bluer as R increases reaching a value of $g-r = 0.15$. This change in the slope of the profile coincides with the radius at which both the galaxy and the disk show a drop in the luminosity-mass ratio (see dashed line in Fig.~\ref{fig:M_vs_L}) and a drop in the age profile (see dashed line in Fig.~\ref{fig:age_TNG}), which would suggest that the outer region of the disk is dominated by a young, luminous, bluer stellar population. This could suggest an inside-out disk formation scenario. However, it should be noted that a more exhaustive analysis of the galaxy formation would be necessary to confirm these hypotheses. The differences in the color profiles between the different methods are negligible.

\section{Conclusions}
\label{sec:conclusion}

We have presented a variation of \citet{abadi_03} dynamic decomposition method, which takes into account the energy distribution in addition to the circularity parameter. We have also implemented this method, together with some other variations in the literature, in a python package (\texttt{GalaxyChop}) that allows to perform the task of dynamical decomposition of galaxies simply, independently of whether it belongs to a cosmological simulation or an isolated equilibrium galaxy model. Additionally, we have analyzed the ability of the 6 different dynamical decomposition methods implemented to recover the stellar components of 9 models of isolated galaxies in equilibrium built with the AGAMA code \citep{agama}. We found that all the methods tested here produce dynamical components with similar global properties, leading to the conclusion that their choice does not dramatically influence the recovered stellar components. However, it is important to note that the choice of the method for performing the dynamical decomposition is decisive for some fundamental properties, such as stellar mass fractions ($f_\mathrm{sph}$ and $f_\mathrm{dsk}$) or scale radii, which are under- or overestimated by up to a factor of two (see upper panel Fig.~\ref{fig:r_half_recuperado}). It should also be noted that the largest differences are found in the inner regions of galaxies, where the spatial intersection of the disk and spheroid makes disentangling the two components more challenging. A detailed analyses of purity and completeness  shows that the JEHistogram method presented here offers substantial improvements in recovering more detailed galaxy features, such as density profiles and velocity profiles, where the task is more difficult and challenging.

As an example of how our code can be applied not only to galaxies in equilibrium but also to galaxies obtained from a cosmological simulation, we apply the 6 decomposition methods to a galaxy from the TNG100 simulation with a stellar mass similar to that of the Milky Way. We identified the disk and spheroid using the \gxchp{} package and we carried out the comparison of some properties. Regarding the mass fractions of the spheroids as well as the stellar mass spatial distribution of the spheroids and disks, we found similar results regardless of the method. This similarity in results is also observed when we compare the half-mass radius and the half-mass height, obtained regardless of the method used. We can again note that the innermost part of the galaxy ($R\lesssim 1.5\ \mathrm{kpc}$) is where it becomes more difficult to carry out the assignment of stellar particles to one component or another. In addition, we compare the profiles obtained by the dynamic decomposition and the classical photometric decomposition, resulting in components that are very different as a consequence of the fact that the mass-luminosity relation varies throughout the galaxy. We find that one should be cautious when comparing dynamic and photometric decomposition. Finally, we find that the dynamically identified disk and spheroid components have different astrophysical properties. The disk has a younger and bluer stellar population than the spheroid. Furthermore, the g-r color profile for the spheroid is approximately constant while the disk shows a change in both the age profile and the color profile, suggesting that the two components underwent different formation processes. This reinforces the idea that the dynamical processes are intrinsically related to the astrophysical processes of the stellar components.

\begin{acknowledgements}

The authors are thankful to the anonymous referee for his comments and suggestions, which helped to greatly improve the first version of this paper. VAC and MA thank the financial support from Agencia Nacional de Promoción Científica y Tecnólogica, the Consejo Nacional de Investigaciones Científicas y Técnicas (CONICET, Argentina) and the Secretaría de Ciencia y Tecnología de la Universidad Nacional de Córdoba (SeCyT-UNC, Argentina).

\end{acknowledgements}
\bibliographystyle{aa}
\bibliography{bibliography.bib}

\begin{thebibliography}{42}
\expandafter\ifx\csname natexlab\endcsname\relax\def\natexlab#1{#1}\fi

\bibitem[{{Abadi} {et~al.}(2003){Abadi}, {Navarro}, {Steinmetz}, \& {Eke}}]{abadi_03}
{Abadi}, M.~G., {Navarro}, J.~F., {Steinmetz}, M., \& {Eke}, V.~R. 2003, \apj, 597, 21

\bibitem[{{Crain} {et~al.}(2015){Crain}, {Schaye}, {Bower}, {Furlong}, {Schaller}, {Theuns}, {Dalla Vecchia}, {Frenk}, {McCarthy}, {Helly}, {Jenkins}, {Rosas-Guevara}, {White}, \& {Trayford}}]{Crain-2015-EAGLE}
{Crain}, R.~A., {Schaye}, J., {Bower}, R.~G., {et~al.} 2015, \mnras, 450, 1937

\bibitem[{{Dalcanton} \& {Bernstein}(2002)}]{2002-Dalcanton}
{Dalcanton}, J.~J. \& {Bernstein}, R.~A. 2002, \aj, 124, 1328

\bibitem[{{de Vaucouleurs}(1959)}]{de_Vaucouleurs_1959}
{de Vaucouleurs}, G. 1959, Handbuch der Physik, 53, 311

\bibitem[{{Dom{\'e}nech-Moral} {et~al.}(2012){Dom{\'e}nech-Moral}, {Mart{\'\i}nez-Serrano}, {Dom{\'\i}nguez-Tenreiro}, \& {Serna}}]{Domenech+12}
{Dom{\'e}nech-Moral}, M., {Mart{\'\i}nez-Serrano}, F.~J., {Dom{\'\i}nguez-Tenreiro}, R., \& {Serna}, A. 2012, \mnras, 421, 2510

\bibitem[{{Dreyer}(1888)}]{Dreyer_1888}
{Dreyer}, J.~L.~E. 1888, \memras, 49, 1

\bibitem[{Du {et~al.}(2019)Du, Ho, Zhao, Shi, Debattista, Hernquist, \& Nelson}]{du_19}
Du, M., Ho, L.~C., Zhao, D., {et~al.} 2019, The Astrophysical Journal, 884, 129

\bibitem[{{Freeman}(1970)}]{Freeman_1970}
{Freeman}, K.~C. 1970, \apj, 160, 811

\bibitem[{{Gargiulo} {et~al.}(2019){Gargiulo}, {Monachesi}, {G{\'o}mez}, {Grand}, {Marinacci}, {Pakmor}, {White}, {Bell}, {Fragkoudi}, \& {Tissera}}]{Gargiulo-2019}
{Gargiulo}, I.~D., {Monachesi}, A., {G{\'o}mez}, F.~A., {et~al.} 2019, \mnras, 489, 5742

\bibitem[{{Herschel}(1864)}]{Herschel_1864}
{Herschel}, J. F.~W. 1864, Philosophical Transactions of the Royal Society of London Series I, 154, 1

\bibitem[{{Hubble}(1926)}]{Hubble_1926}
{Hubble}, E.~P. 1926, \apj, 64, 321

\bibitem[{{Hubble}(1936)}]{Hubble_1936}
{Hubble}, E.~P. 1936, {Realm of the Nebulae} (Yale University Press)

\bibitem[{{Jagvaral} {et~al.}(2022){Jagvaral}, {Campbell}, {Mandelbaum}, \& {Rau}}]{Jagvaral-2022}
{Jagvaral}, Y., {Campbell}, D., {Mandelbaum}, R., \& {Rau}, M.~M. 2022, \mnras, 509, 1764

\bibitem[{{Kormendy} \& {Kennicutt}(2004)}]{2004-Kormendy}
{Kormendy}, J. \& {Kennicutt}, Robert~C., J. 2004, \araa, 42, 603

\bibitem[{{Lindblad}(1933)}]{Lindblad_1933}
{Lindblad}, B. 1933, Handbuch der Astrophysik, 5, 937

\bibitem[{{Marinacci} {et~al.}(2014){Marinacci}, {Pakmor}, \& {Springel}}]{2014_Marinacci}
{Marinacci}, F., {Pakmor}, R., \& {Springel}, V. 2014, \mnras, 437, 1750

\bibitem[{{Marinacci} {et~al.}(2018){Marinacci}, {Vogelsberger}, {Pakmor}, {Torrey}, {Springel}, {Hernquist}, {Nelson}, {Weinberger}, {Pillepich}, {Naiman}, \& {Genel}}]{Marinacci-2018-TNG}
{Marinacci}, F., {Vogelsberger}, M., {Pakmor}, R., {et~al.} 2018, \mnras, 480, 5113

\bibitem[{{Messier}(1781)}]{Messier_1781}
{Messier}, C. 1781, {Catalogue des N{\'e}buleuses et des Amas d'{\'E}toiles (Catalog of Nebulae and Star Clusters)}, Connoissance des Temps ou des Mouvements C{\'e}lestes

\bibitem[{{Naiman} {et~al.}(2018){Naiman}, {Pillepich}, {Springel}, {Ramirez-Ruiz}, {Torrey}, {Vogelsberger}, {Pakmor}, {Nelson}, {Marinacci}, {Hernquist}, {Weinberger}, \& {Genel}}]{Naiman-2018-TNG}
{Naiman}, J.~P., {Pillepich}, A., {Springel}, V., {et~al.} 2018, \mnras, 477, 1206

\bibitem[{{Nelson} {et~al.}(2018){Nelson}, {Pillepich}, {Springel}, {Weinberger}, {Hernquist}, {Pakmor}, {Genel}, {Torrey}, {Vogelsberger}, {Kauffmann}, {Marinacci}, \& {Naiman}}]{Nelson-2018-TNG}
{Nelson}, D., {Pillepich}, A., {Springel}, V., {et~al.} 2018, \mnras, 475, 624

\bibitem[{{Obreja} {et~al.}(2018){Obreja}, {Macci{\`o}}, {Moster}, {Dutton}, {Buck}, {Stinson}, \& {Wang}}]{obreja_gsf_code}
{Obreja}, A., {Macci{\`o}}, A.~V., {Moster}, B., {et~al.} 2018, \mnras, 477, 4915

\bibitem[{{Obreja} {et~al.}(2016){Obreja}, {Stinson}, {Dutton}, {Macci{\`o}}, {Wang}, \& {Kang}}]{Obreja_2016}
{Obreja}, A., {Stinson}, G.~S., {Dutton}, A.~A., {et~al.} 2016, \mnras, 459, 467

\bibitem[{{Okamoto} {et~al.}(2008){Okamoto}, {Nemmen}, \& {Bower}}]{Okamoto-2008}
{Okamoto}, T., {Nemmen}, R.~S., \& {Bower}, R.~G. 2008, \mnras, 385, 161

\bibitem[{{Park} {et~al.}(2019){Park}, {Yi}, {Dubois}, {Pichon}, {Kimm}, {Devriendt}, {Choi}, {Volonteri}, {Kaviraj}, \& {Peirani}}]{2019_Park}
{Park}, M.-J., {Yi}, S.~K., {Dubois}, Y., {et~al.} 2019, \apj, 883, 25

\bibitem[{Pedregosa {et~al.}(2011)Pedregosa, Varoquaux, Gramfort, Michel, Thirion, Grisel, Blondel, Prettenhofer, Weiss, Dubourg, Vanderplas, Passos, Cournapeau, Brucher, Perrot, \& Duchesnay}]{scikit-learn}
Pedregosa, F., Varoquaux, G., Gramfort, A., {et~al.} 2011, Journal of Machine Learning Research, 12, 2825

\bibitem[{{Pillepich} {et~al.}(2018){Pillepich}, {Nelson}, {Hernquist}, {Springel}, {Pakmor}, {Torrey}, {Weinberger}, {Genel}, {Naiman}, {Marinacci}, \& {Vogelsberger}}]{Pillepich-2018-TNG}
{Pillepich}, A., {Nelson}, D., {Hernquist}, L., {et~al.} 2018, \mnras, 475, 648

\bibitem[{{Sandage} \& {Tammann}(1981)}]{Sandage&Tammann_1981}
{Sandage}, A. \& {Tammann}, G.~A. 1981, {A Revised Shapley-Ames Catalog of Bright Galaxies}

\bibitem[{{Scannapieco} {et~al.}(2009){Scannapieco}, {White}, {Springel}, \& {Tissera}}]{Scannapieco-2009}
{Scannapieco}, C., {White}, S. D.~M., {Springel}, V., \& {Tissera}, P.~B. 2009, \mnras, 396, 696

\bibitem[{{Schaye} {et~al.}(2015){Schaye}, {Crain}, {Bower}, {Furlong}, {Schaller}, {Theuns}, {Dalla Vecchia}, {Frenk}, {McCarthy}, {Helly}, {Jenkins}, {Rosas-Guevara}, {White}, {Baes}, {Booth}, {Camps}, {Navarro}, {Qu}, {Rahmati}, {Sawala}, {Thomas}, \& {Trayford}}]{Schaye-2015-EAGLE}
{Schaye}, J., {Crain}, R.~A., {Bower}, R.~G., {et~al.} 2015, \mnras, 446, 521

\bibitem[{{Schwarz}(1978)}]{1978_Schwarz}
{Schwarz}, G. 1978, Annals of Statistics, 6, 461

\bibitem[{{S{\'e}rsic}(1963)}]{sersic_1963}
{S{\'e}rsic}, J.~L. 1963, Boletin de la Asociacion Argentina de Astronomia La Plata Argentina, 6, 41

\bibitem[{{Springel} {et~al.}(2018){Springel}, {Pakmor}, {Pillepich}, {Weinberger}, {Nelson}, {Hernquist}, {Vogelsberger}, {Genel}, {Torrey}, {Marinacci}, \& {Naiman}}]{Springel-2018-TNG}
{Springel}, V., {Pakmor}, R., {Pillepich}, A., {et~al.} 2018, \mnras, 475, 676

\bibitem[{{Tissera} {et~al.}(2012){Tissera}, {White}, \& {Scannapieco}}]{2012_Tissera}
{Tissera}, P.~B., {White}, S. D.~M., \& {Scannapieco}, C. 2012, \mnras, 420, 255

\bibitem[{{Trujillo} \& {Bakos}(2013)}]{2013-Trujillo}
{Trujillo}, I. \& {Bakos}, J. 2013, \mnras, 431, 1121

\bibitem[{van~den Bosch {et~al.}(1999)van~den Bosch, Lewis, Lake, \& Stadel}]{van_den_Bosch_1999}
van~den Bosch, F.~C., Lewis, G.~F., Lake, G., \& Stadel, J. 1999, The Astrophysical Journal, 515, 50

\bibitem[{{Vasiliev}(2018)}]{AGAMA-docu}
{Vasiliev}, E. 2018, arXiv e-prints, arXiv:1802.08255

\bibitem[{{Vasiliev}(2019)}]{agama}
{Vasiliev}, E. 2019, \mnras, 482, 1525

\bibitem[{{Vogelsberger} {et~al.}(2014){Vogelsberger}, {Genel}, {Springel}, {Torrey}, {Sijacki}, {Xu}, {Snyder}, {Nelson}, \& {Hernquist}}]{2014_Vogelsberger}
{Vogelsberger}, M., {Genel}, S., {Springel}, V., {et~al.} 2014, \mnras, 444, 1518

\bibitem[{{Vogelsberger} {et~al.}(2020){Vogelsberger}, {Marinacci}, {Torrey}, \& {Puchwein}}]{2020_Vogelsberger}
{Vogelsberger}, M., {Marinacci}, F., {Torrey}, P., \& {Puchwein}, E. 2020, Nature Reviews Physics, 2, 42

\bibitem[{{Xu} {et~al.}(2019){Xu}, {Zhu}, {Grand}, {Springel}, {Mao}, {van de Ven}, {Lu}, {Wang}, {Pillepich}, {Genel}, {Nelson}, {Rodriguez-Gomez}, {Pakmor}, {Weinberger}, {Marinacci}, {Vogelsberger}, {Torrey}, {Naiman}, \& {Hernquist}}]{Xu-2019}
{Xu}, D., {Zhu}, L., {Grand}, R., {et~al.} 2019, \mnras, 489, 842

\bibitem[{{Yu} {et~al.}(2023){Yu}, {Bullock}, {Gurvich}, {Hafen}, {Stern}, {Boylan-Kolchin}, {Faucher-Gigu{\`e}re}, {Wetzel}, {Hopkins}, \& {Moreno}}]{Yu-2023}
{Yu}, S., {Bullock}, J.~S., {Gurvich}, A.~B., {et~al.} 2023, \mnras, 523, 6220

\bibitem[{{Zana} {et~al.}(2022){Zana}, {Lupi}, {Bonetti}, {Dotti}, {Rosas-Guevara}, {Izquierdo-Villalba}, {Bonoli}, {Hernquist}, \& {Nelson}}]{Zana-2022}
{Zana}, T., {Lupi}, A., {Bonetti}, M., {et~al.} 2022, \mnras, 515, 1524

\end{thebibliography}

\begin{appendix}
    
\section{Application example of the package}
\label{ap:tutorial}

As an example of the application of the package, we show how to reach to the dynamical decomposition for a simulated galaxy using the \texttt{JHistogram} method.
For this example we used a galaxy selected from TNG100 simulation \citep{Pillepich-2018-TNG, Naiman-2018-TNG, Nelson-2018-TNG, Marinacci-2018-TNG, Springel-2018-TNG}. This dataset is available in \texttt{GalaxyChop} repository for the user to test the functionalities. 
The models return the star particles and their {\it label} indicating to which component belongs. Depending on the model, it is the number of components in which the galaxy is decomposed.
    
\begin{lstlisting}
# Import necessary packages
>>> import numpy as np
>>> import astropy.units as u

# Import GalaxyChop package
>>> import galaxychop as gchop

# Loading the data
>>> gal = gchop.read_hdf5("galaxy.h5")

# Galaxy centring and alignment
>>> galc = gchop.center(gal)
>>> gal = gchop.star_align(galc, r_cut=30)

# Dynamical decomposition models
# JHistogram Model (Abadi+03)
>>> decomposer = gchop.models.JHistogram()
>>> components = decomposer.decompose(gal)
>>> components.describe()
         Particles           Deterministic mass         
              Size  Fraction               Size Fraction
Spheroid     12785  0.342918       1.223594e+10  0.32792
Disk         24498  0.657082       2.507786e+10  0.67208
    
# Index of the component each stellar
# particles belongs to.
>>> labels= 
... components.labels[components.ptypes == 'stars']
>>> print(label)
array([0. 1. 0. ... 0. 0. 0.])

# Index = 0: correspond to spheroid particles.
# Index = 1: correspond to disc particles.  
\end{lstlisting}
    
A more detailed tutorial of \texttt{GalaxyChop} applications can be found in the repository\footnote{\url{https://galaxy-chop.readthedocs.io/en/latest/}}.

\end{appendix}

\end{document}